%% file: IEEE-main.tex
\documentclass[conference]{IEEEtran}
\IEEEoverridecommandlockouts
\usepackage{framed}

\usepackage{cite}
\usepackage{graphicx}
\usepackage{textcomp}
\usepackage{xcolor}
\usepackage{subfigure}
\usepackage{listings}
\usepackage{xcolor,pifont}
\usepackage{url}
\usepackage[colorlinks,urlcolor=blue, linkcolor=black]{hyperref}
\usepackage{balance}
\usepackage{xspace}
\usepackage{bbding}
\usepackage{xcolor}
\usepackage{longtable}
\usepackage{booktabs}
\usepackage{array}
\usepackage{algorithm}
\usepackage{paralist}
\usepackage{multirow,makecell}
\usepackage{enumerate}
\usepackage{fancynum}

\newcolumntype{L}[1]{>{\raggedright\let\newline\\\arraybackslash\hspace{0pt}}m{#1}}
\newcolumntype{C}[1]{>{\centering\let\newline\\\arraybackslash\hspace{0pt}}m{#1}}
\newcolumntype{R}[1]{>{\raggedleft\let\newline\\\arraybackslash\hspace{0pt}}m{#1}}

\usepackage{adjustbox}
\newcolumntype{R}[2]{%
    >{\adjustbox{angle=#1,lap=\width-(#2)}\bgroup}%
    l%
    <{\egroup}%
}

\renewcommand{\texttt}[1]{%
 \begingroup
 \ttfamily
 \begingroup\lccode`~=`/\lowercase{\endgroup\def~}{/\discretionary{}{}{}}%
 \begingroup\lccode`~=`[\lowercase{\endgroup\def~}{[\discretionary{}{}{}}%
 \begingroup\lccode`~=`.\lowercase{\endgroup\def~}{.\discretionary{}{}{}}%
 \catcode`/=\active\catcode`[=\active\catcode`.=\active
 \scantokens{#1\noexpand}%
 \endgroup
 }
 
\def\BibTeX{{\rm B\kern-.05em{\sc i\kern-.025em b}\kern-.08em
    T\kern-.1667em\lower.7ex\hbox{E}\kern-.125emX}}
\begin{document}

\title{Don't Fish in Troubled Waters! Characterizing Coronavirus-themed Cryptocurrency Scams}

\author{
\IEEEauthorblockN{Pengcheng Xia$^{1}$, Haoyu Wang$^{1}$, Xiapu Luo$^{2}$, Lei Wu$^{3}$, Yajin Zhou$^{3}$, Guangdong Bai$^{4}$,\\Guoai Xu$^{1}$, Gang Huang$^{5}$, Xuanzhe Liu$^{5}$}
\IEEEauthorblockA{
$^{1}$ Beijing University of Posts and Telecommunications, Beijing, China
}
\IEEEauthorblockA{
$^{2}$ The Hong Kong Polytechnic University
$^{3}$ Zhejiang University
}
\IEEEauthorblockA{
$^{4}$ The University of Queensland
$^{5}$ Peking University
}
}

\maketitle

\begin{abstract}
As COVID-19 has been spreading across the world since early 2020, a growing number of malicious campaigns are capitalizing the topic of COVID-19. COVID-19 themed cryptocurrency scams are increasingly popular during the pandemic. However, these newly emerging scams are poorly understood by our community. In this paper, we present the first measurement study of COVID-19 themed cryptocurrency scams. We first create a comprehensive taxonomy of COVID-19 scams by manually analyzing the existing scams reported by users from online resources. Then, we propose a hybrid approach to perform the investigation by: 1) collecting reported scams in the wild; and 2) detecting undisclosed ones based on information collected from suspicious entities (e.g., domains, tweets, etc). We have collected 195 confirmed COVID-19 cryptocurrency scams in total, including 91 token scams, 19 giveaway scams, 9 blackmail scams, 14 crypto malware scams, 9 Ponzi scheme scams, and 53 donation scams. We then identified over 200 blockchain addresses associated with these scams, which lead to at least 330K US dollars in losses from 6,329 victims.
For each type of scams, we further investigated the tricks and social engineering techniques they used. To facilitate future research, we have released all the well-labelled scams to the research community.

\end{abstract}

\input{intro.tex}

\input{background.tex}

\input{CryptoScams.tex}

\input{Measurement.tex}
\input{Discussion.tex}

\newpage

\balance
\bibliographystyle{IEEEtran}
\bibliography{ref.bib}

\end{document}

%% file: intro.tex
\section{Introduction}
The coronavirus pandemic (COVID-19) has quickly become a world-wide crisis.
Beyond the virus in the physical world, the cyber-space also suffers from the security threats relevant to COVID-19.
Recent months have witnessed a surge of malicious campaigns that are exploiting the pandemic, such as email scams, ransomware, phishing domains and malicious apps~\cite{covidtreat,covidattack,he2020beyond,apwgreport}, etc.
According to the report from Federal Trade Commission (FTC), victims in the US have lost over \$77 million to fraud during this crisis by the time of July 2020~\cite{FTCScamLoss}, and the number is just the `tip of the iceberg', as the fraud was typically under-reported by consumers.

Blockchain, as one of the most popular techniques in recent years, has attracted great attentions from attackers in this pandemic. 
As more and more businesses accept cryptocurrencies as payments and people have been drawn to cryptocurrencies, more scammers have appeared to take advantage of these eager new targets to steal money.
According to an FBI press released on April 2020, the number of scams related to cryptocurrency has increased greatly during the COVID-19 pandemic~\cite{fbipress}. 
For example, some scammers posing as World Health Organization (WHO) sent fake emails asking for Bitcoin donation~\cite{fakewho}. They also used a forged email address, ``donate@who.int'', to defraud people. Also, it is reported that a malicious COVID-19 themed domain  \texttt{coronavirusapp.site} claims to offer a real-time coronavirus tracking app~\cite{covidlock}. However, the app is a new kind of ransomware called ``CovidLock'' that locks the victim's devices and requests for Bitcoins in 48 hours on the ransom note. 
Besides, a number of Initial Coin Offering (ICO) and other token scam projects are taking advantage of COVID-19 to release trashy cryptocurrency tokens (e.g., CoronaCoin, COVID19 Coin), to cheat inexperienced investigators~\cite{coronacoin}.
For example, three consecutive \textit{exit scams} happened for the CoronaCoin, which broke the project.

\textbf{This Work.}
In this paper, we take the first step to characterize the coronavirus-themed cryptocurrency scams. 
Our goal is to systematically \textit{summarize and investigate different types of cryptocurrency scams related to COVID-19}, \textit{explain how they work}, \textit{measure their prevalence} and \textit{characterize their impacts}. 
To this end, we first make efforts to create a taxonomy of COVID-19 themed cryptocurrency scams (see \textbf{Section~\ref{sec:taxonomy}}). 
By resorting to security reports of COVID-19 cybersecurity attacks and scam reports obtained from discussion forums, we have summarized a taxonomy of 6 types of scams that take advantage of both COVID-19 and cryptocurrency to infect unsuspecting users. 
These scams include: 1) COVID-19 token scam, 2) COVID-19 giveaway scam, 3) COVID-19 blackmail scam, 4) crypto malware scam, 5) COVID-19 themed Ponzi scheme, and 6) crypto donation scam. 
By demystifying these types of scams, we propose a hybrid approach to collect scams in the wild (see \textbf{Section~\ref{sec:detection}}), and identify 195 COVID-19 themed cryptocurrency scams in total, which correlate with 201 scam blockchain addresses, 57 scam domains, 14 crypto malware, 47 social accounts and 91 coronavirus-related tokens.
We further measure the characteristics and impacts of these scams (see \textbf{Section~\ref{sec:measurement}}). 
Our investigation shows that at least 330K US dollars have been stolen by the attackers from $6,329$ victims, which is a lower bound estimate of the prevalence and criminal profits associated with these scams.

%To the best of our knowledge, this paper gives a first impression on the landscape of the COVID-19 themed crypto scams, revealing some unexpected and interesting observations.
To the best of our knowledge, this paper takes the first step to reveal the COVID-19 themed cryptocurrency scams with some unexpected and interesting observations.
We believe this work shall shed some light on identifying scams related to public events. To boost future research, we have released all the collected COVID-19 cryptocurrency scams to the community at: \url{https://covid19scam.github.io}.

%\begin{center}
%\url{https://covid19scam.github.io}
%\end{center}

%% file: background.tex
\section{Background and Related Work}
\label{sec:background}

\subsection{Blockchain and Cryptocurrency}
Blockchain was invented in 2008 to act as a public, decentralised ledger for Bitcoin. It stores transactions or related events among involved parties. Each transaction in a block is verified by consensus on most of the system's participants, and the data stored in the blockchain cannot be modified. 
Cryptocurrency is a kind of digital asset that uses cryptography to ensure the security of its creations and transactions. Since the debut of Bitcoin, thousands of cryptocurrencies are emerging~\cite{allcrypto}. Cryptocurrencies except Bitcoin can be classified into two types: \textit{Altcoins}, which mean alternatives to Bitcoin, and \textit{tokens}, which are unable to operate independently without existing blockchain platforms.
Blockchains like Ethereum and EOSIO, have simplified the development of token smart contracts. One can create a token smart contract with just a few lines of code~\cite{TokenScope}. Recent work~\cite{tokenWWW} suggested that there are over 160,000 tokens exist on Ethereum.

\subsection{Cryptocurrency Scams}
More and more attackers target on cryptocurrencies to make a profit. 
%As cryptocurrencies are gaining more and more attention, malicious actors are also devoted to making profit on this rich field. 
In 2019, cryptocurrency scams caused over \$ 4.26 billion in losses~\cite{cryptoscam}. Although users, wallets and exchanges are adopting new countermeasures to avoid being scammed, new scam techniques still emerge to defraud users' money. Vasek and Moore surveyed the presence of Bitcoin-based scams~\cite{vasek2015there} in 2015. By gathering reports from voluntary vigilantes and reports tracked in online forums, they identified 192 scams and categorized them into four groups: \textit{Ponzi schemes}, \textit{mining scams}, \textit{scam wallets} and \textit{fraudulent exchanges}. 
Most of existing scam studies were focused on detecting the Ponzi schemes~\cite{chen2018detecting, bartoletti2020dissecting, bartoletti2018data,vasek2018analyzing,chen2019detecting,toyoda2017identification,toyoda2019novel}, fraudulent Initial Coin Offering (ICO)~\cite{liebau2019crypto,zetzsche2017ico}, market manipulation of cryptocurrencies~\cite{gandal2018price,chen2019market,hamrick2018economics,chen2019detecting,hamrick2018examination}, blockchain honeypots~\cite{torres2019art}, and phishing scams~\cite{wu2019phishers,phillips2020tracing}, etc. 
To the best of our knowledge, this paper is the first work to study COVID-19 themed cryptocurrency scams.

\iffalse
\subsection{Security Analysis of Blockchain Ecosystem}
In order to better deal with attacks on blockchain, many researchers have studied the blockchain ecosystem on multiple aspects. On consensus mechanism, Bissas et al.~\cite{bissias2016analysis} presented a mathematical model of mining process and use it to evaluate the double-spend attack. Huang et al.~\cite{EOSIO} studied the activities on the EOSIO and found over 30\% of the accounts in the platform are bots and 301 attack accounts in the real world. Chen et al.~\cite{chen2018understanding} conducted a systematic study on Ethereum by graph analysis and proposed approaches to address attack forensics and anomaly detection. Atzei et al.~\cite{atzei2017survey} studied attacks on Ethereum smart contracts and summarized the  major vulnerabilities on Solidity, EVM bytecode, and blockchain levels. Cryptocurrency exchanges have also been studied by many researchers~\cite{kim2018risk,mccorry2018preventing,chohan2018problems,feder2017impact,moore2018revisiting}. For example, Feder et al.~\cite{feder2017impact} investigated the impact of distributed denial-of-service (DDoS) attacks and other disruptions on Bitcoin ecosystem, and found that the number of large trades on the Mt. Gox exchange fell sharply after the DDoS attacks. McCorry et al.~\cite{mccorry2018preventing} proposed a reactive mechanism to detect heist targeting at cryptocurrency exchanges and freeze all withdraws. This countermeasure also allows an exchange to bring a trusted vault key online to recover from compromise.
\fi

\subsection{COVID-19 related Research}
Since its outbreak, coronavirus has attracted great attentions from various research communities. A large number of studies were focused on the medical domain, including pathology study, epidemiology study, treatment study and so on~\cite{bai2020presumed,zu2020coronavirus,onder2020case,shen2020treatment}. 
There are also some sociology or 
psychological studies on COVID-19 like misinformation research or social impact analysis~\cite{pennycook2020fighting,nelson2020sociological}. Computer scientists have adopted computing techniques like machine learning to help medical practitioners deal with COVID-19. Zhang et al.~\cite{zhang2020covid} proposed the confidence-aware anomaly detection (CAAD) model to screen viral pneumonia on chest X-ray images. Wang et al.~\cite{wang2020covid} proposed COVID-Net, a deep convolutional neural network design tailored for the detection of COVID-19 cases from chest X-ray (CXR) images.

%cybersecurity regarding these illegal behaviors . Some studies on them have been conducted
At the same time, however, cyber attackers also exploit this situation to conduct criminal activities~\cite{lallie2020cyber,mathewcybersecurity,radanliev2020digitalization,kallberg2020covid,ahmad2020corona,he2020beyond}. Lallie et al.~\cite{lallie2020cyber} analyzed the timeline of COVID-19 themedcyber-attacks.
%%and found that there is a large gap between initial outbreak and first COVID-19 related cyber-attack. Furthermore, 
They utilised the UK as a case study to demonstrate how cyber-criminals leveraged key events and governmental announcements to carefully craft and design attacks. 
Ahmad et al.~\cite{ahmad2020corona} analyzed the security challenges on the work-from-home model and proposed some recommendations on how to avoid being attacked by phishing websites or emails. 
He et al.~\cite{he2020beyond} presented a systematic analysis of coronavirus-themed malware and found that over 53\% of the malicious apps they collected were camouflaged as official apps. They also found that these apps aim to steal users' private information or to make profit by phishing and extortion.
Sun et al.~\cite{sun2020vetting} analyzed the security and privacy issues of COVID-19 contact tracing apps.
Although a few previous studies mentioned COVID-19 themed scams, there lacks a systematic study on COVID-19 themed cryptocurrency scams, such as the impact and popularity of these scams.

%% file: CryptoScams.tex
\section{A Taxonomy of COVID-19 Crypto Scams}
\label{sec:taxonomy}

To understand the prevalence and characteristics of COVID-19 themed crypto scams in the wild, we first create a comprehensive taxonomy by manually analyzing the existing scam reports.
We will explain how each kind of scams works in this section.
Based on this taxonomy, we will further measure their popularity and impacts in Section~\ref{sec:detection} and Section~\ref{sec:measurement}.

\begin{table*}[t]
\centering
\caption{A Taxonomy of 6 Kinds of COVID-19 Cryptocurrency Scams.}
\resizebox{\linewidth}{!}{
\begin{tabular}{|c|c|c|c|}
\hline
Type of   Scams &
  Tricks &
  Scam Entities &
  Example \\ \hline
COVID-19 token scam &
  \begin{tabular}[c]{@{}c@{}}Issuing new tokens that claim to be used for charity; \\  ``Pump-and-dump'' schemes
  \end{tabular} &
  Token &
  \begin{tabular}[c]{@{}c@{}}CoronaCoin \\ (0x10Ef64cb79Fd4d75d4Aa7e8502d95C42124e434b)\end{tabular} \\ \hline
COVID-19 giveaway scam &
  \begin{tabular}[c]{@{}c@{}}Promising to reward users based on the money they sent.\end{tabular} &
  \begin{tabular}[c]{@{}c@{}}Social Network\\Domain \end{tabular} &
  https://gatesbtc.live \\ \hline
COVID-19 themed Blackmail &
  \begin{tabular}[c]{@{}c@{}}Threatening to spread coronavirus if not receive cryptos, \\ or selling products/cures for COVID-19.\end{tabular} &
  Email &
  abtconsultnl@oceanenergy.ch \\ \hline
Crypto Malware &
  \begin{tabular}[c]{@{}c@{}}Locking victims' phones or computers \\ and asking for cryptos\end{tabular} &
  Software &
  https://coronavirusapp.site \\ \hline
Ponzi Schemes &
  \begin{tabular}[c]{@{}c@{}}Claiming to return high interest payback \\ if someone invest with cryptos\end{tabular} &
  \begin{tabular}[c]{@{}c@{}}Domain, \\ Social Network\end{tabular} &
  https://coronainvest.io \\ \hline
Donation Scams &
  \begin{tabular}[c]{@{}c@{}}Acting as  public organizations and \\ claiming to raise money for COVID-19\end{tabular} &
  \begin{tabular}[c]{@{}c@{}}Domain, \\ Social Network\end{tabular} &
  https://covid-coin.com \\ \hline
\end{tabular}
}
\label{tab:scamintro}
\end{table*}

\textbf{Creating the Taxonomy.}
We resort to the following information to create the taxonomy:
1) security reports of COVID-19 cybersecurity attacks released by security companies (e.g., DomainTools~\cite{domaintools} and McAfee~\cite{mcafee}), 
2) over 40 scam reports obtained from scam accusations on Bitcointalk~\cite{bitcointalk} and BitcoinAbuse~\cite{bitcoinabuse}, 
3) COVID-19 themed scams summarized by the Federal Trade Commission (FTC)~\cite{FTCBlog}, 
4) COVID-19 scams reported in the threat intelligence platforms (e.g., AlienVault~\cite{alienvault} and COVID-19 MISP~\cite{misp}), and 
5) the related information by searching keywords like `COVID-19 attack' and `COVID-19 scam' on Google.
We summarize a taxonomy of 6 types of scams related to COVID-19, as shown in Table~\ref{tab:scamintro}. We will detail these scams in the following subsections.

\begin{figure}[t]
\centering
\includegraphics[width=0.45\textwidth]{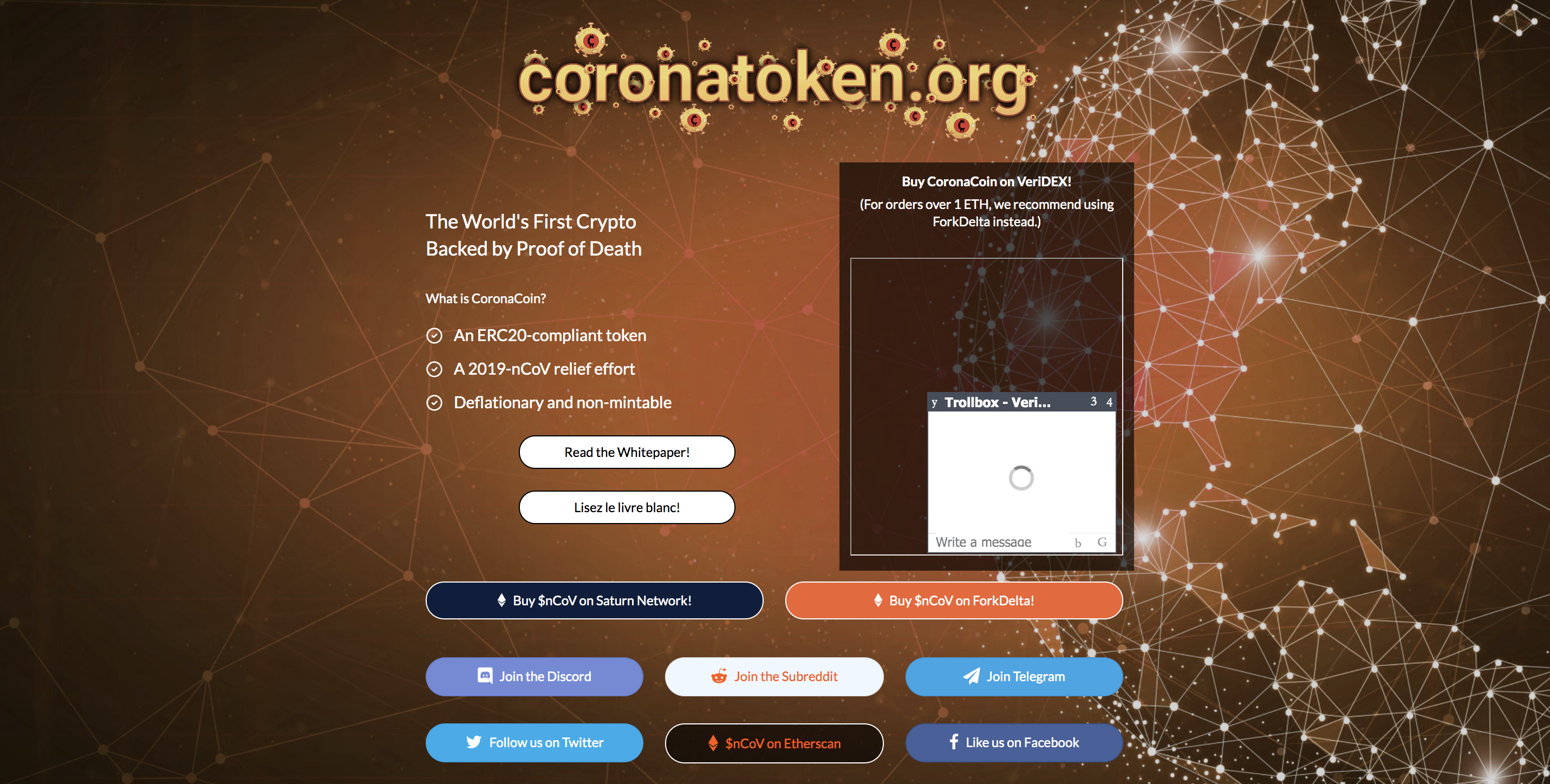}
\caption{Screenshot of \textit{coronatoken.org} (the official website of CoronaCoin).}
\vspace{-0.1in}
\label{fig:coronatoken}
\end{figure}

\subsection{COVID-19 Token Scam}
\label{taxonomy:token}

A number of Initial Coin Offering (ICO) projects and scam token projects take advantage of COVID-19 to release trashy cryptocurrency tokens to cheat inexperienced investigators.
In addition to claiming to help people alleviate the pain of virus and lockdown, the founding teams of some tokens also indicate that they will participate in public welfare activities and donate part of the profit of the tokens to charitable organizations. 
Moreover, many tokens can be traded on some decentralized cryptocurrency exchanges (DEX) and their promotion tricks make investors believe these tokens are profitable. However, some tokens are totally scams since the very beginning, and the founders would just disappear with the investments they received from the ICO projects. Some token projects may first run normally after issuing the tokens, but the project owners who possess most of the tokens will monetize their tokens when the price increases. This type of scams is also known as the ``Pump-and-dump Scheme''~\cite{pumpscam}, which has been studied extensively by many researchers~\cite{hamrick2018economics,chen2019detecting}. 

Figure~\ref{fig:coronatoken} shows the official webpage of the CoronaCoin\footnote{Ethereum address: 0x10Ef64cb79Fd4d75d4Aa7e8502d95C42124\-e434b} (Token symbol: NCOV), which is claimed to be the first token related to COVID-19. It has received more and more attention since reported by medias like Reuters, Nasdaq and New York Times~\cite{coronacoin1}. 
It is designed to be burned per 48 hours according to the number of infections and casualties from the Coronavirus. The founding team calls this mechanism ``proof of death''. 
It is advertised that by this way, the token will be deflationary and its value will increase. Although they donated about \$235 on March 6th to the Red Cross for the first time, which seems to prove they operated this project with a good will, three consecutive \textit{exit scams} happened afterwards. 
The CoronaCoin's developers and administrators monetized the large amount of tokens they managed based on the ``Pump-and-dump Schemes''. These scams broke the project, leading to its current failure. From the scam accusation reports on BitcoinTalk~\cite{coronacointalk}, many investors have found that this token was a scam at the very beginning and they condemn the behaviors of designing scams by exploiting the pandemic, but there are still many unsuspecting investors who were deceived. 

\subsection{COVID-19 Giveaway Scam}
\label{taxonomy:giveaway}

Giveaway scam is a type of commonly used trick in the field of cryptocurrency scam, and it is no exception when it comes to the COVID-19 themed scams. 
The malicious actors promise to reward users based on the money (tokens) that the users send, but they will not fulfill their promises in the end. 
The giveaway scam can be delivered using both social network (e.g., Twitter) and content sharing services (e.g., YouTube).

\begin{figure}[t]
\centering
\subfigure[A YouTube video that promotes \textit{http://gatesbtc.live} (a domain involved in COVID-19 themed giveaway scams).]{
\begin{minipage}[t]{0.95\linewidth}
\centering
\includegraphics[width=0.95\textwidth]{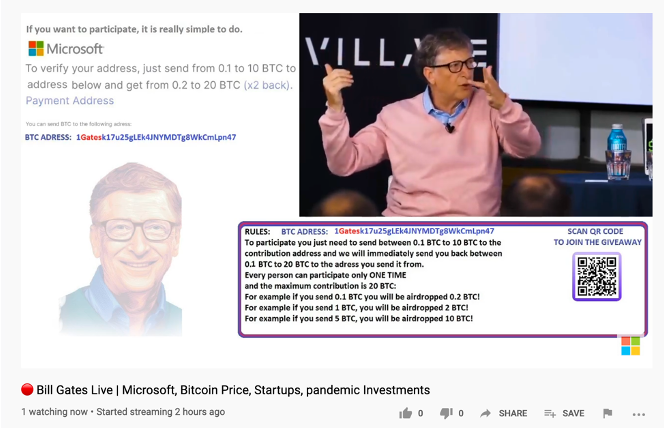}
\end{minipage}
}

\subfigure[A tweet posted by the hacked Twitter account \textit{Warren Buffett}.]{
\begin{minipage}[t]{0.95\linewidth}
\centering
\includegraphics[width=0.95\textwidth]{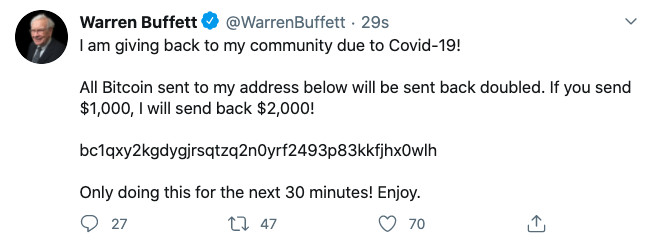}
\end{minipage}
}
\caption{Two examples of COVID-19 themed giveaway scams.}
\label{fig:giveaway}
\end{figure}

Figure~\ref{fig:giveaway}(a) shows an example of giveaway scam reported on BitcoinAbuse. 
This YouTube video shows Bill Gates' speech about Bitcoin and pandemic investment, which inserts a giveaway scam Bitcoin address 1Gatesk17u25gLEk4JNYMDTg8WkCmLpn47. 
It asks users to send Bitcoins to this address so that they will gain a double payback.
%%The video asks for bitcoins to send to this address and promises a double reward bitcoins based how many they receive.
It is interesting to see that there is a ``Gates'' in the address, which may increase the credibility of the scam. 
%%By the time of this study, this address has received about 0.13 BTC in 9 times.

Giveaway scams are also prevalent on social network.
Note that, our investigation reveals that the scams can be distributed by either the \textit{scam social network accounts} or the \textit{hacked accounts}.
For the scam social network accounts, they are usually fake accounts that act as the famous people. For example, there are many fake accounts that have the same name, avatar and other information with Vitalik Buterin (one of the co-founders of Ethereum). In this way, they can post giveaway scam tweets to cheat unsuspecting users.
For the hacked accounts, one of the largest campaigns is the Twitter's massive attack on July 15th, 2020~\cite{twitterhack}. The twitter accounts of major companies and individuals were compromised and these accounts were controlled to conduct a COVID-19 themed giveaway scam. Figure~\ref{fig:giveaway}(b) shows one of the scam tweets. The attackers used Warren Buffett's account to ask for sending Bitcoins to their addresses and promised to return a double payback. Till July 16th, the scam address bc1qxy2kgdygjrsqtzq2n0yrf2493p83kkfjhx0wlh has received $12.02$ BTC (about 110K US dollars\footnote{As the prices fluctuate all the time, we estimate the prices of cryptocurrencies based on their closing prices on July 16th (the same below).}). 

%We also find some tweets on Twitter that claims to make some coronavirus-related giveaway events for those who follow, like, retweet and tag their friends. Though we cannot get their scam address due to their complex scam process, we assume they are scams based on their past tweets and their newly registration time. There is an example on Figure~\ref{fig:giveaway} (b). The Twitter account was registered on April this year, and most of his tweets are about giveaway events, making us think it is a scam account.

\begin{figure}[h]
\centering
\includegraphics[width=0.45\textwidth]{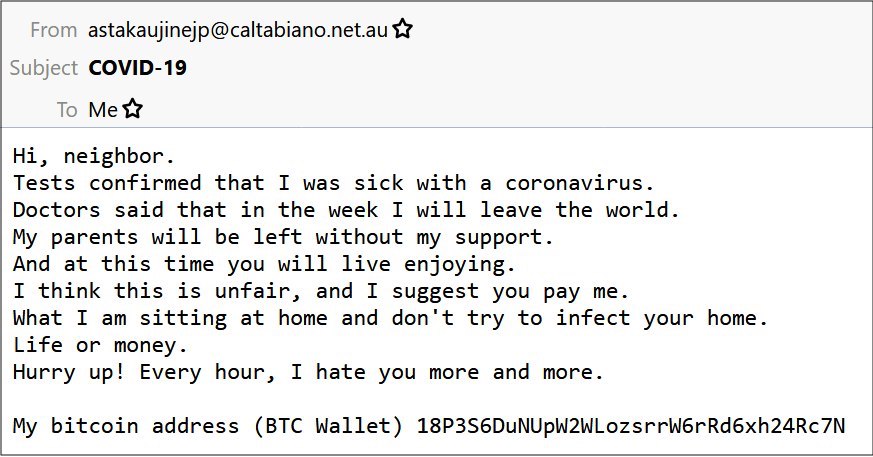}
\caption{An example of a COVID-19 crypto extortion email.}
\vspace{-0.1in}
\label{fig:extortion}
\end{figure}

\subsection{COVID-19 themed Blackmail}
\label{taxonomy:mail}

The cryptocurrency-related extortion emails are mainly focused on sextortion, according to previous work~\cite{SpamsSextortion}. 
During this pandemic, attackers have adopted some other social engineering techniques for extortion. For example, attackers can ask for money by threatening that they are able to infect the emails' receivers with coronavirus.
There are also some scam emails claiming that they have a cure for coronavirus. 

Figure~\ref{fig:extortion} shows an example of a COVID-19 themed crypto extortion email.
The attacker claims to be the victim's neighbour that was infected with COVID-19. He asks the receiver to send Bitcoins to 18P3S6DuNUpW2\-WLozsrrW6rRd\-6xh24Rc7N, otherwise he will spread the virus. This address has been reported 15 times on BitcoinAbuse by the time of this study. Fortunately, it only receives $0.06$ US dollars in total. We will further discuss the social engineering techniques used in these scams and their impacts in Section~\ref{sec:measurement}.

\begin{figure}[h]
\centering
\includegraphics[width=0.45\textwidth]{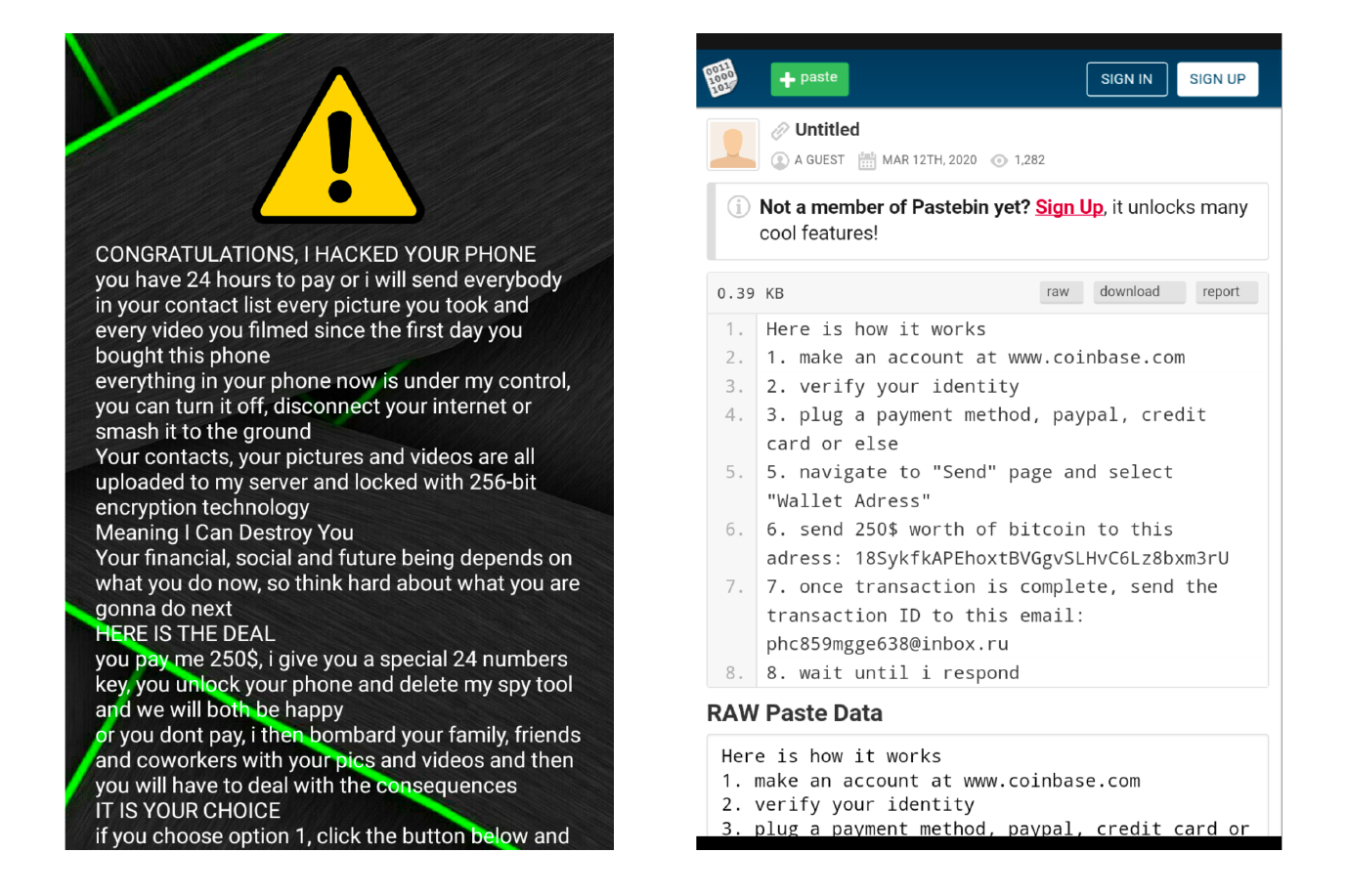}
\vspace{-0.1in}
\caption{An example of a COVID-19 themed ransomware.}
\vspace{-0.1in}
\label{fig:ransom}
\end{figure}

\subsection{COVID-19 Crypto Malware}
\label{taxonomy:ransomware}

Ransomware is a type of crypto malware that threatens to publish the victim's data or perpetually block access to it unless a ransom is paid. Ransomware is quite popular in recent years, which has been widely studied by both cybersecurity industry and academia~\cite{ransomstate,paquet2019ransomware,homayoun2019drthis}. 
Ransomware attack is typically carried out using a Trojan that is disguised as a legitimate application. COVID-19 has been invoked for malicious purposes by the ransomware creators. 
They usually impersonate COVID-19 themed apps to trick users into installing them. Once launched, it will encrypt the victim's mobile phone files or force a lock screen and extort a high ransom. Besides, there are also other crypto malware like crypto miners in the wild.

Figure~\ref{fig:ransom} shows an example of a mobile ransomware\footnote{MD5:d1d417235616e4a05096319bb4875f57}, which was distributed on a malicious COVID-19 themed domain \textit{http://coronavirusapp.site}. Once launched, the malware will lock user's phone and ask for Bitcoins. Clicking the button on the screen will redirect users to a Pastebin link (https://pastebin.com/GK8qrfaC) where records the attacker's Bitcoin address \textit{18SykfkAPEhoxtBVGgvSLHvC6Lz8bxm3rU}.

\subsection{Covid-19 themed Ponzi scheme}
\label{taxonomy:ponzi}

Ponzi scheme is a kind of scam promising high rates of return with little risk to investors~\cite{ponzidefine}.
In fact, the attackers will only pay back the early investors using the money that the subsequent investors deposit, and they will finally take all the money they get and leave. Lots of studies were focused on this kind of scam, both on normal blockchain Ponzi schemes~\cite{bartoletti2018data} and Ponzi schemes built on smart contracts~\cite{chen2018detecting,bartoletti2020dissecting}. Malicious actors also exploit the pandemic to carry out Ponzi schemes, claiming to help people reduce economic pressure of income fluctuations during quarantine. 

\begin{figure}[t]
\centering
\includegraphics[width=0.45\textwidth]{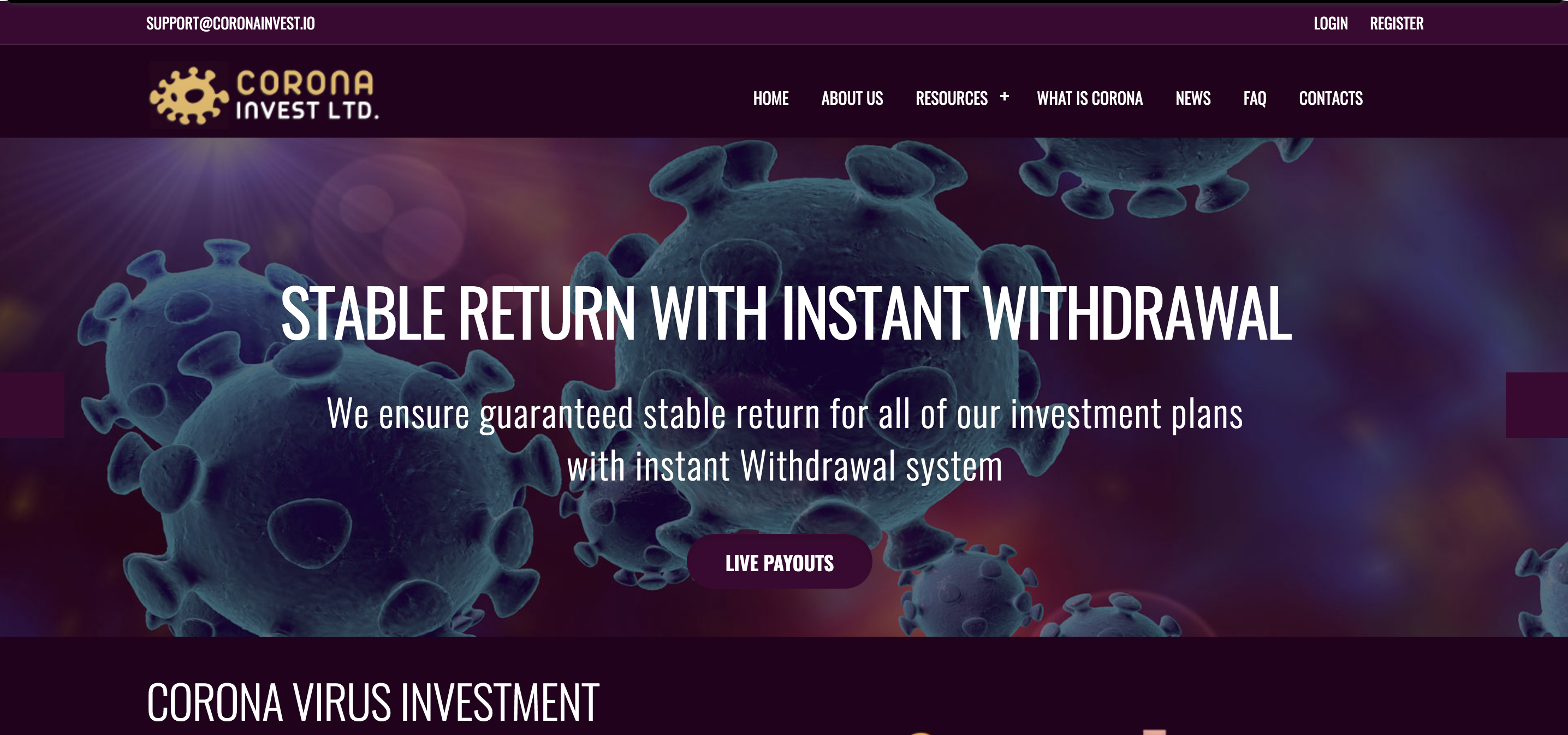}
\caption{Screenshot of \textit{http://coronainvest.io} (a Ponzi scheme domain.)}
\label{fig:ponzi}
\end{figure}

Figure~\ref{fig:ponzi} shows an example of a Ponzi scheme domain. 
The attackers claimed that they will invest in coronavirus' vaccine and give stable and high return to the investors. With such a promise, the website did not survive for a long time. According to Wayback Machine\footnote{http://web.archive.org/web/20200315040815/https://coronainvest.io/}, the website can be accessed earlier than March 15th and it was shut down by the time of March 29th.

\begin{figure}[h]
\centering
\includegraphics[width=0.45\textwidth]{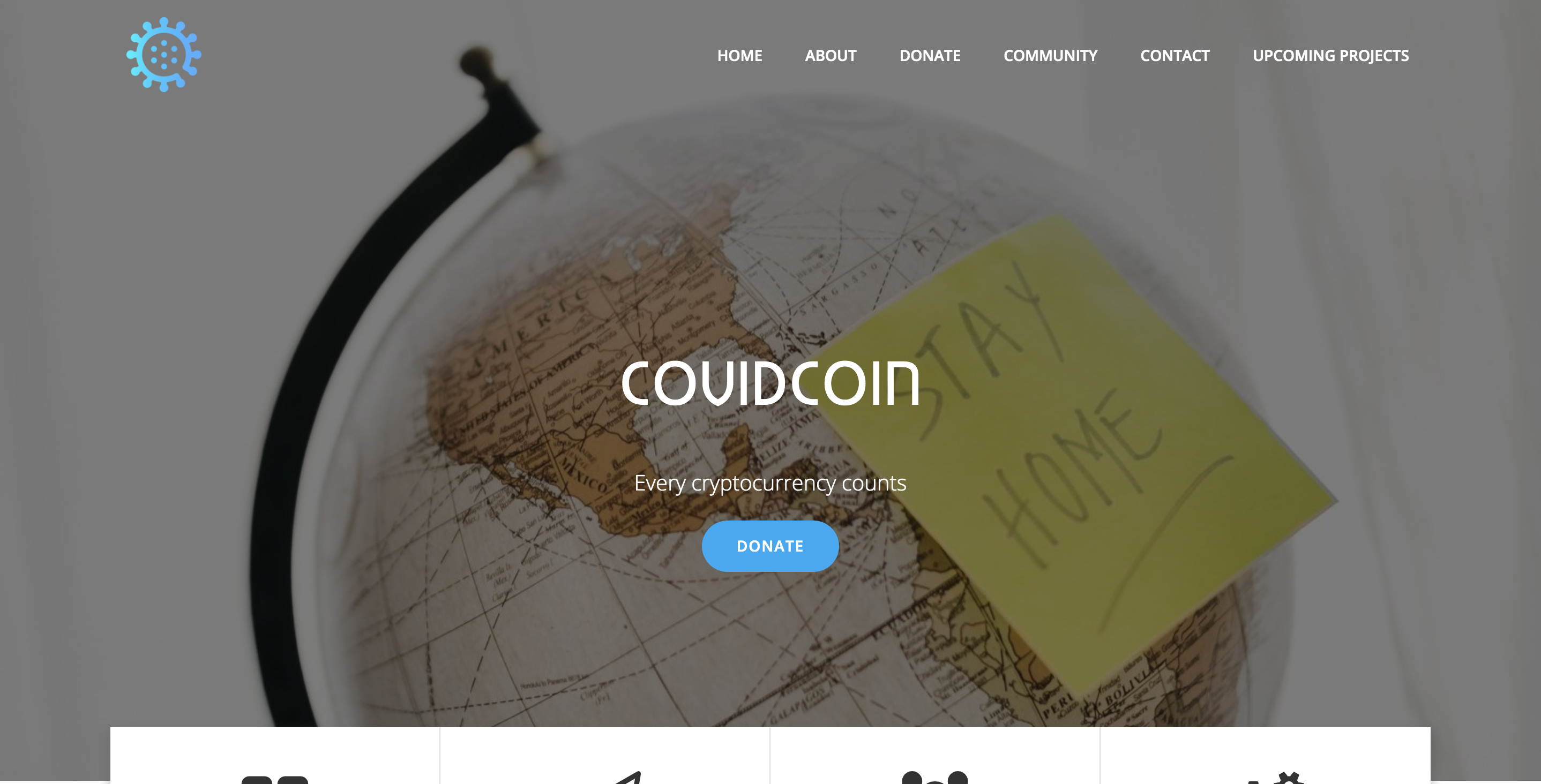}
\caption{Screenshot of \textit{http://covid-coin.com/} (a donation scam domain).}
\label{fig:donation}
\end{figure}

\begin{figure}[h]
\centering
\includegraphics[width=0.48\textwidth]{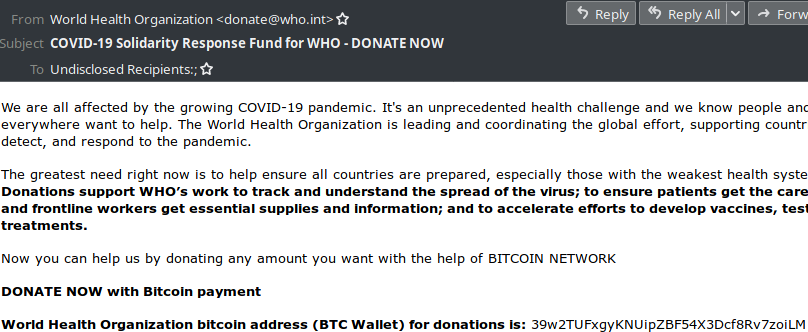}
\caption{Screenshot of an Email sent from ``donate@who.int'' (donation scam).}
\label{fig:donationemail}
\end{figure}

\subsection{Fake Crypto Donation}
\label{taxonomy:donation}

Besides some traditional scams on cryptocurrency, we also identify a number of coronavirus-themed crypto donation scams spread via emails, social network and websites. 
Attackers may act as some health-related official organizations or departments like Centers for Disease Control and Prevention (CDC), World Health Organization (WHO) and the United Nations International Children's Fund (UNICEF).
Moreover, some of them act as a charity group or individual to ask for help during this pandemic.
In general, they will ask people to send money to their cryptocurrency wallet addresses in the name of donation.

Figure~\ref{fig:donation} shows a fake donation domain, which was flagged as malicious by VirusTotal. The attackers call for the cryptocurrency community to fight against the virus by donating to them. 
The domain is built on top of WordPress, which supports donations from 7 kinds of cryptocurrencies. 
As another example, Figure~\ref{fig:donationemail} shows a donation scam in which the attacker pretended to be WHO (using the email address ``donate@who.int''). Its sole purpose is to ask for donations to COVID-19 Solidarity Response Fund.

%% file: Measurement.tex
\begin{figure}[ht]
\centering
\includegraphics[width=0.48\textwidth]{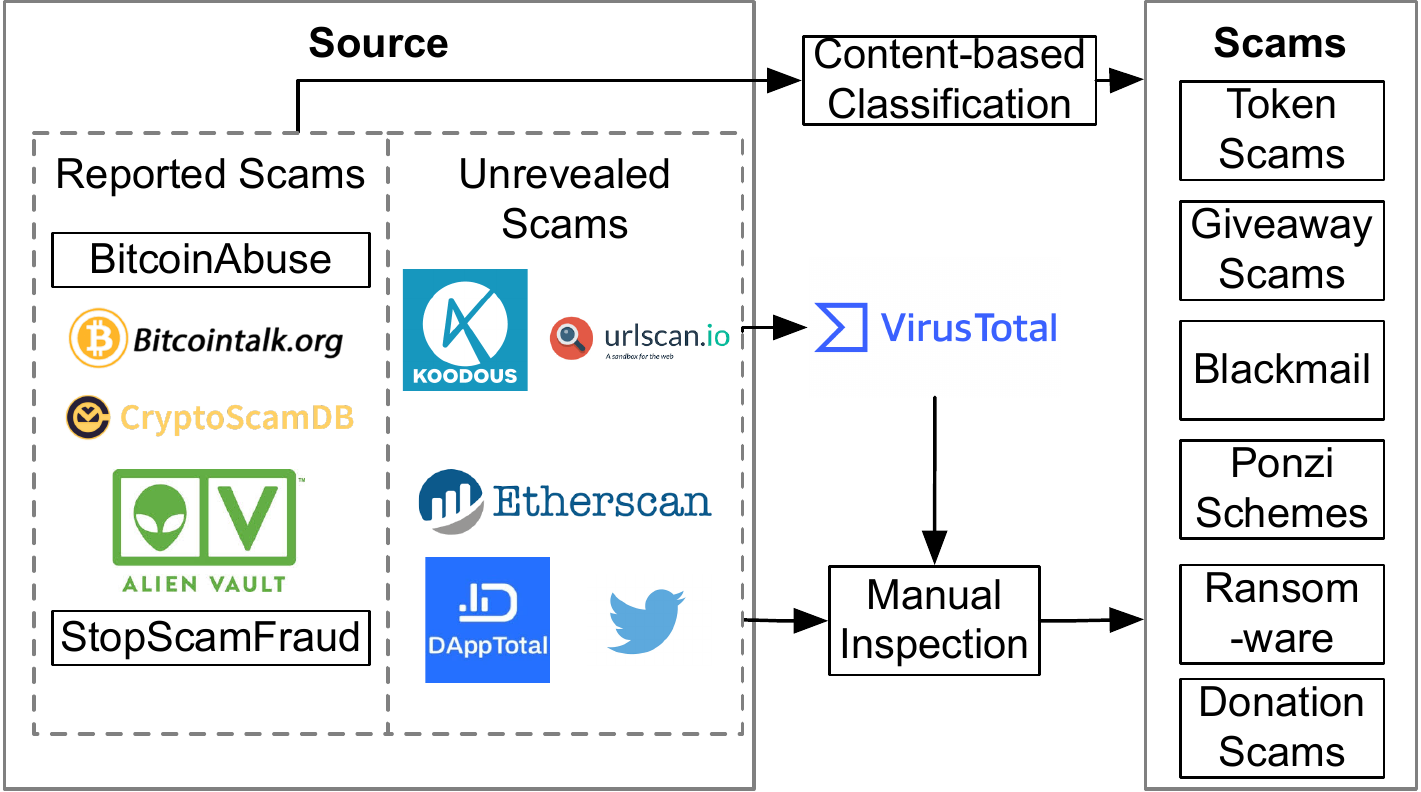}
\caption{Our approach to collecting COVID-19 themed cryptocurrency scams.}
\vspace{-0.1in}
\label{fig:datasetworkflow}
\end{figure}

\section{Collecting COVID-19 Themed Crypto Scams}
\label{sec:detection}

In order to fully understand the prevalence of COVID-19 scams in the wild, we have adopted a hybrid approach to conduct the investigation by: 1) collecting reported scams in the wild; and 2) detecting undisclosed ones based on information collected from various sources. 
Figure~\ref{fig:datasetworkflow} shows the overall procedure of our approach.

\subsection{Harvesting Reported Scams}

Based on the taxonomy of COVID-19 scams summarized in Section~\ref{sec:taxonomy}, we resort to the following scam databases for collecting known ones reported by users.

\textbf{1) BitcoinAbuse.} 
BitcoinAbuse~\cite{bitcoinabuse} is a database where users can report malicious or scam Bitcoin addresses they encountered. By the time of this study, BitcoinAbuse has aggregated more than 170K Bitcoin addresses, which are mainly used in ransomware, blackmails, and giveaway scams. To identify the COVID-19 themed scams, we use keywords like ``COVID'' or ``Corona'' to fetch related scam reports. Through this, 31 COVID-19 themed scam Bitcoin addresses are identified from BitcoinAbuse. Based on the tags and descriptions provided by users, we manually classify them into 17 donation scams, 9 blackmail scams, and 5 giveaway scams.
    
\textbf{2) CryptoScamDB.}
CryptoScamDB~\cite{cryptoscamdb} aims at collecting malicious cryptocurrency domains and related blockchain addresses using a crowd-sourcing based method. 
By the time of this study, there are over 7,700 malicious cryptocurrency domains collected by CryptoScamDB.
To identify COVID-19 related scams, we use the heuristics to search keywords like ``COVID'' or ``corona'' in both the domain names and the labels provided by CryptoScamDB. However, we did not find any scams related to COVID-19 from CryptoScamDB.
    
\textbf{3) BitcoinTalk.}
BitcoinTalk~\cite{bitcointalk} is an online forum devoted to the discussion of Bitcoin and other cryptocurrencies. It hosts a ``Scam Accusations'' board for people to report scams. Thus, we have implemented a crawler to get all the related posts, and then identify scams related to COVID-19 using keywords matching. Finally, we get 9 related posts. We then manually analyzed the archived pages and corresponding user reports to label them. At last, we have identified 4 token scams, 2 Ponzi schemes, 2 ransomware and 1 donation scam from these posts.
    
\textbf{4) Threat Intelligence Platforms.} 
Threat intelligence platforms like AlienVault~\cite{alienvault} and some security companies like McAfee provide reports related to coronavirus-themed attacks. These reports often contain indicators of compromise (IoCs), which can reflect attackers' intentions to a certain extent. We use a crawler to fetch all the reports and identify 9 COVID-19 cryptocurrency scam reports (with 11 blockchain addresses). Based on the detailed descriptions provided by these reports, we have categorized them into 4 donation scams, 5 crypto ransomware and 2 giveaway scams.
    
\textbf{5) StopScamFraud.} StopScamFraud~\cite{stopscamfraud} is a platform devoted to collecting users' reports on scam emails. By the time of this study, there are over 900 mail accusations related to COVID-19. As we only consider cryptocurrency related scams, we search the scams using blockchain related keywords (e.g., Bitcoin, Ethereum, etc.). However, we do not find any related scam accusations.

%\textbf{Scam Databases.}

% We first resort to scam databases like BitcoinAbuse and CryptoScamDB\footnote{https://cryptoscamdb.org/} to collect reported scams related to COVID-19. 
% These two databases are actively maintained and can often bring up-to-date scams reported by users. 
% We use keywords like ``COVID'' or ``Corona'' to fetch coronavirus-related reports or domains on them. We end up finding 32 reports on them. We fetch all the information includes tags, descriptons and addresses of these reports, and classify them based on their tags and descriptions. Thus, 32 reports are divided into 17 donation scams, 10 blackmails, 5 giveaway scams.  

% \textbf{Crypto Forums.}
% Since cryptocurrency scams' victims may also go to crypto forums for help or accusation, a crawler is implemented to fetch the scam accusations on the crypto forums like BitcoinTalk. Thus, 11 posts are found related to coronavirus-related scams. Based on the descrption of each post, we identify 6 token scams, 2 Ponzi schemes, 2 ransomware and 1 donation scam from these posts. 

% \textbf{Threat Intelligence Platforms.}
% Besides, we design a crawler for threat intelligence platforms like AlienVault and reports from cybersecurity companies like McAfee to search for indicator of compromise (IoCs) realted to COVID-19. Because not all the IoCs are related to cryptocurrency scams, after getting all the IoCs, we filter the data we collect based on their tags or descriptions. We get 9 scam cases in this step, including 2 donation scams, 5 ransomware and 2 giveaway scams. 

\begin{table*}[ht]
\centering
\caption{The distribution of scams harvested from different sources.}
\vspace{-0.1in}
\begin{tabular}{@{}|c|c|cc|cccc|c|@{}}
\toprule
Type of Scams &
  \begin{tabular}[c]{@{}c@{}}Total \\ Found\end{tabular} &
  \begin{tabular}[c]{@{}c@{}}Reported \\ Scams\end{tabular} &
  \begin{tabular}[c]{@{}c@{}}Unrevealed \\ Scams\end{tabular} &
  \begin{tabular}[c]{@{}c@{}}Related \\ Tokens\end{tabular} &
  \begin{tabular}[c]{@{}c@{}}Related \\ Domains\end{tabular} &
  \begin{tabular}[c]{@{}c@{}}Related \\ Binaries\end{tabular} &
  \begin{tabular}[c]{@{}c@{}}Related \\ Social \\ Accounts\end{tabular} &
  \begin{tabular}[c]{@{}c@{}}Extracted \\ Addresses\end{tabular} \\ \midrule
Token Scams    & 91           & 4           & 87           & 91          & 14          & 0           & 7           & 91           \\
Giveaway Scams & 19           & 7           & 12           & 0           & 5           & 0           & 10          & 21           \\
Blackmails     & 9           & 9          & 0            & 0           & 0           & 0           & 0           & 9           \\
Crypto Malware & 14           & 5           & 9            & 0           & 6           & 14          & 0           & 4            \\
Ponzi Schemes  & 9            & 2           & 7            & 0           & 9           & 0           & 0           & 0            \\
Donation Scams & 53           & 21          & 32           & 0           & 23          & 0           & 30          & 76           \\ \midrule
\textbf{Total} & \textbf{195} & \textbf{48} & \textbf{147} & \textbf{91} & \textbf{57} & \textbf{14} & \textbf{47} & \textbf{201} \\ \bottomrule
\end{tabular}
\vspace{-0.1in}
\label{tab:scamtotal}
\end{table*}

\subsection{Detecting Unrevealed Scams}

To further identify unrevealed COVID-19 themed cryptocurrency scams in the wild, we first perform a semi-automated analysis to identify suspicious scam entities, and then manually verify them. As summarized in Table~\ref{tab:scamintro}, the scams can be delivered via entities including tokens, domain, malware, smart contract (DApp), email, and social network. Thus, our goal is to first identify the scam entities related to COVID-19.

\textbf{1) Scam Tokens.}
Etherscan~\cite{etherscan} is an Ethereum blockchain explorer that tracks tokens on Ethereum. We resort to Etherscan to search for COVID-19 themed ERC-20 and ERC-721 tokens\footnote{ERC-20 and ERC-721 are both token standards in Ethereum. ERC is a set of rules that the developers have to follow so they can implement a token in the Ethereum blockchain ecosystem. It includes information about the protocol specifications and the description of the contract.} using keywords like ``corona'' and ``COVID-19''.
By the time of our study, we have identified 87 tokens that target the pandemic. To verify whether they are indeed scams, on one hand, we search these tokens (with their corresponding addresses) in Google to check whether they have related websites, social accounts or scam reports. On the other hand, we manually analyzed all their smart contracts, to see if they use simple ERC-20 or ERC-721 code templates without additional functions. Based on our manually verification, we believe all of them are trashy tokens without any value. COVID-19 is only used as the publicity stunt to cheat unsuspecting users.
    
\textbf{2) Scam Domains.}
We seek to identify the malicious domains related to COVID-19 first, and then try to detect whether some of them belong to the scams we summarized. Here, we take advantage of URLScan~\cite{urlscan}, an online service that provides history snapshots, IP resolutions and other detailed information of massive domains. 
We use a number of keywords including ``coronavirus'', ``COVID-19'' and their squatting ones (e.g., cor`a'navirus, cor`oo'navirus)\footnote{Previous work suggested that typosquatting~\cite{typosquatting} and combosquatting~\cite{combosquatting} attacks are prevalent in malicious domains.} to identify the domain names that contain the keywords from URLScan. Combining with COVID-19 themed domains fed by RiskIQ~\cite{riskiq}, we have identified $175,966$ newly registered domains that contain COVID-19 related keywords since January 2020. 
    
Note that, the COVID-19 themed domains reported by these threat intelligence platforms are not necessary to be malicious, as some of them were found via newly registered domain feeds, which never served any malicious activities. Thus, we further take advantage of VirusTotal~\cite{virustotal}, a widely-used online service that aggregates over 60 anti-virus engines, to check whether the COVID-19 domains are malicious. At last, we identify $101,004$ COVID-19 domains flagged as malicious by at least one anti-virus engine on VirusTotal.
For the malicious domains, after excluding the domain parking web pages, we use a heuristic approach to identify and classify the cryptocurrency related scams. 
Since domains that contain words like ``crypto'', ``bitcoin'', ``invest'', ``donation'' tend to carry out coronavirus-themed cryptocurrency scams, we select such domains that contain keywords in their domain names, as the scam candidates (over 150 domains found).
Then, we perform a manually inspection to determine whether they are cryptocurrency scams. For example, if a malicious domain claims that it will give high excess returns to the investors (i.e., high-yield investment program), we will label it as a Ponzi scheme. 
If a malicious domain camouflages as reputable organizations to ask for donation, we will label it as a donation scam.
If a domain claims to reward users based on the cryptocurrency that they transferred, we will label the domain as a giveaway scam and further label its corresponding blockchain address as a scam address. At last, we have manually verified 7 Ponzi schemes, 2 donation scams, and 2 giveaway scams in this way.

    %%Moreover, we fetch all the files that are downloaded from these domains or communicate to these domains through VirusTotal. Thus, we get $31,542$ malicious domains and $5,954$ apps. 
\textbf{3) Malware.}
To identify the COVID-19 themed crytocurrency malware, we rely on two data sources. On one hand, for the aforementioned $101,004$ COVID-19 domains flagged as malicious, we use a premium API provided by ViusTotal to get files related to these malicious domains, i.e., files that communicate to these domains or files that are downloaded from these domains. After this step, we have collected $7,294$ binaries and $2,362$ of them are flagged as malicious by at least one engine on VirusTotal. On the other hand, we use Koodous~\cite{koodous}, a large Android app repository with over 62 million apps in total by the time of this work, to find suspicious COVID-19 themed malware. Koodous contains historical apps from various sources. In this work, we collect the apps whose app names or package names contain COVID-19 related keywords. In this way, $2,378$ apps are collected. For all the collected binaries, we further use VirusTotal and AVClass~\cite{AVClass} for labelling their malware families. Note that, not all the samples can be assigned families, as AVClass needs a number of AV engines' detection results to achieve an agreement on family name. As we only consider COVID-19 themed ransomware and cryptomining malware in this work, all the malware whose families are labelled ``Ransomware'', ``Locker'' or ``Coinminer'' are kept to the further research. Thus, 4 COVID-19 ransomware and 5 COVID-19 cryptomining malware are identified. All of them were labelled by at least 6 anti-virus engines, and 5 of them were flagged by at least 30 anti-virus engines.
    
\textbf{4) Scam Social Network Posts/Accounts.}
Social network is one of the major channels for distributing and advertising COVID-19 themed scams. Thus, we resort to Twitter and Telegram to identify more scams. To be specific, we first identify Tweets and Telegram discussions that contain both the COVID-19 keywords and cryptocurrency keywords. Then, we manually inspect their contents. For example, if we find a Twitter account that imitates an official account of reputable organizations to publish donation information, we will flag it as a donation scam. If we identify Tweets that advertise giveaway information, we will mark it as a giveaway scam.
We have analyzed all the related Tweets and Telegram discussions from January 2020 to July 2020, and identified 30 donation scams and 10 giveaway scams, which were distributed by 37 Twitter accounts and 3 Telegram accounts.
    
\textbf{5) Scam Smart Contracts (DApps).}
Previous work~\cite{chen2019detecting,bartoletti2020dissecting} suggested that some DApps (smart contracts) are actually Ponzi schemes. 
To verify whether the Ponzi DApps take advantage of COVID-19, we resort to DAppTotal~\cite{dapptotal}, a well-known DApp explorer to find whether there are coronavirus-themed DApps. However, our keywords searching does not find any DApps related to COVID-19.

Note that, as we cannot get unrevealed email scams from public information, we did not identify any new scam emails besides the reported ones. For all the detected scams, we further analyze them using a semi-automated approach to identify their correlated blockchain addresses (if available).
To be specific, we first use regular expressions\footnote{For example, the regular expressions (bc1|[13])[a-zA-HJ-NP-Z0-9]{25,39} is used to identify Bitcoin address.} to identify blockchain address candidates, and then we manually verify whether they are real addresses.

\subsection{Dataset Overview}

Table~\ref{tab:scamtotal} shows the statistics of scams we collected. 
\textbf{Overall, we have identified 195 COVID-19 cryptocurrency scams}, including 91 token scams, 19 giveaway scams, 9 blackmails, 14 crypto malware, 9 Ponzi schemes and 53 donation scams.
These scams correspond to 91 COVID-19 tokens, 57 malicious COVID-19 domains, 14 COVID-19 themed malware, and 47 social accounts on Twitter and Telegram.
Besides, we have identified 201 blockchain addresses correlated with them. 
%%Note that, in our dataset, there are 8 giveaway social accounts that only have posts of ``retweet, like and tag'' contents, 8 software that are not downloadable from VirusTotal and 9 Ponzi schemes that have only snapshots of their homepages. Thus, for these 25 scams, we cannot collected their blockchain addresses.

% \subsection{Identification and Classification of Collected Scams}
% \textbf{Classification of Confirmed Scams.}

% \textbf{Identification and Classification of Suspicious Scams.}

\section{Measurement of COVID-19 Crypto Scams}
\label{sec:measurement}

In this section, we analyze the overall trends of COVID-19 crypto scams, investigate the tricks and social engineering techniques used in scams, and further measure the impacts.

\subsection{The Trends of COVID-19 Cryptocurrency Scams}

%%\haoyu{distribution, evolution, created time, overall statistics like transactions, connected addresses, money flow, etc. Note that, this subsection only shows the overall trends.}

%%\haoyu{we should have a table to show the overall impacts (money) caused by the scams, and list each category, and then depict them in the following.}

\subsubsection{Distribution of Scams} 
The distribution of scams is shown in Table~\ref{tab:scamoverall}. Obviously, the \textit{token scams} (46\%) and \textit{donation scams} (27\%) are dominant in the ecosystem.
There might be two reasons. 
First, cryptocurrencies are claimed to be the \textit{Safe Haven Asset} (compared with the fiat currency). Thus, during the pandemic, attackers have the motivation to use COVID-19 token scams to cheat inexperienced users.
Second, most people want to help fight the pandemic, so the scammers take advantage of this opportunity to act as the official agencies (e.g., WHO) to advertise a number of donation scams.

\subsubsection{Overall Impacts of Scams}
We further estimate the overall impacts of scams.
It is non-trivial to estimate the impacts, as we can only resort to the blockchain transactions related to these scam addresses, which is a lower-bound estimation.
We have collected all the transaction records related to these scam addresses till July 16th 2020.
Since a few scam addresses have been active for a long time before 2020, we only consider the transactions related to these scam addresses since the beginning of the pandemic (early 2020) when calculating the overall financial losses of scams.
Note that token scam is a special case. Here, we only consider the holders who have the corresponding COVID-19 trashy tokens. 
As there are 91 scam tokens, we estimate their prices based on the latest value shown in exchanges. For the tokens we cannot get their latest value from exchanges (which means that they have no trade records on exchanges), we do not count their value.
Thus, \textit{our estimation is definitely a lower bound of the COVID-19 cryptocurrency scam ecosystem}.

As shown in Table~\ref{tab:scamoverall} (column 4 and column 5), the overall number of financial losses is over 333K US dollars, contributed by $6,329$ victims. 
\textit{Giveaway scam} is the most profitable category (over \$287K), with over 500 victims were scammed. 
Besides, at least \$21K contributed by 103 victims were received by \textit{donation scams}, and the volume of \textit{token scams} is over \$23K. We will further investigate each scam category in the following subsections.

% \usepackage{booktabs}
\begin{table}[t]
\centering
\caption{The distribution of 6 types of scams.}
\begin{tabular}{@{}cccrr@{}}
\toprule
Category &
  \begin{tabular}[c]{@{}c@{}}\# of \\ Scam Cases\end{tabular} &
  \begin{tabular}[c]{@{}c@{}}\# of \\ Addresses\end{tabular} &
  \begin{tabular}[c]{@{}c@{}}Est. \\ Victims\end{tabular} &
  \begin{tabular}[c]{@{}c@{}}Est. Scammed \\ Money (\$)\end{tabular} \\ \midrule
Token Scams          & 91           & 91           & 5,701           & 23,178.0                 \\
Giveaway Scams       & 19           & 21           & 516          & 287,663.5         \\
Donation Scams       & 53           & 76           & 103          & 21,788.9         \\
Blackmails     & 9           & 9           & 6            & 514.5         \\
Ransomware     & 14           & 4            & 3            & 102.1         \\
Ponzi Schemes  & 9            & 0            & 0            & 0                  \\ \midrule
\textbf{Total} & \textbf{195} & \textbf{201} & \textbf{6,329} & \textbf{333,247.0} \\ \bottomrule
\end{tabular}
\vspace{-0.1in}
\label{tab:scamoverall}
\end{table}

% \begin{figure}[t]
% \centering
% \includegraphics[width=0.45\textwidth]{images/4_0_1-scamdistribution.pdf}
% \caption{The distribution of 6 kinds of scams.}
% \label{fig:scamdistribution}
% \end{figure}

\subsubsection{The Distribution of Scam Blockchain Addresses} 
\label{subsec:distribution}
Table~\ref{tab:alladdr} shows the distribution of the 201 scam addresses\footnote{Note that we list token scams' addresses separately, because these scams are based on comprehensive designs like ``Pump-and-dump'' schemes and these addresses can not be simply measured with transactions on addresses.}. 
It can be observed that, 
Bitcoin addresses dominate the scams (40.3\%), followed by Ethereum with 18 addresses. 
Among these scam addresses, 138 of them have transactions in 2020.

\begin{table}[t]
\centering
\caption{The 8 Types of Cryptocurrency Addresses in the Dataset.}
\resizebox{\linewidth}{!}{
\begin{tabular}{@{}ccccc@{}}
\toprule
\begin{tabular}[c]{@{}c@{}}Coin or \\ Token\end{tabular} &
  \begin{tabular}[c]{@{}c@{}}\# of Addresses \\ (\# of Addresses Active in 2020)\end{tabular} &
  \begin{tabular}[c]{@{}c@{}}Incoming \\ Transactions\\ in 2020\end{tabular} &
  \begin{tabular}[c]{@{}c@{}}Total \\ Received\\ in 2020\end{tabular} &
  \begin{tabular}[c]{@{}c@{}}Current \\ Value(\$)\end{tabular} \\ \midrule
BTC            & 81 (38)           & 550          & 33.2      & 303,100.5           \\
ETH            & 18 (6)            & 73           & 13.7      & 3,218.2             \\
BCH            & 5 (0)             & 0            & 0.0       & 0.0                \\
LTC            & 2 (1)             & 1            & 2.0       & 84.1               \\
TRX            & 1 (1)             & 2            & 211,254.3  & 3,621.7             \\
USDT           & 1 (1)             & 1            & 44.5      & 44.5               \\
DOGE           & 1 (0)             & 0            & 0.0       & 0.0                \\
XRP            & 1 (0)             & 0            & 0.0       & 0.0                \\ 
Scam Tokens    & 91 (91)           &      -        &     -      & 23,178.0
          \\
\midrule
\textbf{Total} & \textbf{201 (138)} & \textbf{627} & \textbf{-} & \textbf{333,247.0} \\ \bottomrule
\end{tabular}
}
\label{tab:alladdr}
\end{table}

\begin{figure}[t]
\centering
\includegraphics[width=0.45\textwidth]{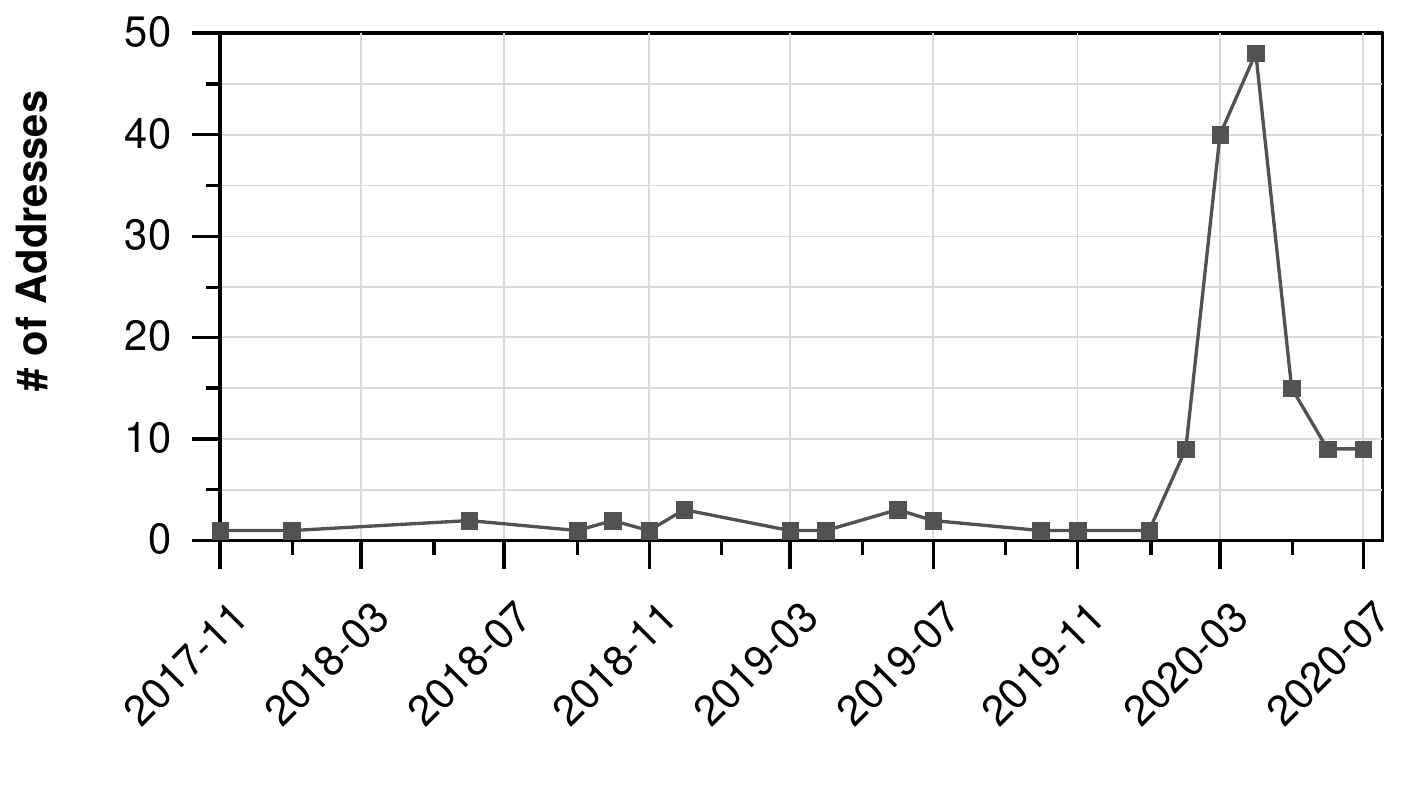}
\vspace{-0.1in}
\caption{The distribution of the first transaction time of scam addresses.}
\label{fig:addrtime}
\end{figure}

\textbf{The Evolution of Scam Addresses.} 
We further analyze the evolution of all the 151 scam addresses that ever have transaction records (including 13 addresses only received transactions before 2020), as shown in Figure~\ref{fig:addrtime}. 
It is interesting to see that, there are 20 addresses in total that were active before 2020. We manually analyzed the 20 addresses, and observed that 2 of them were involved in known scams in 2018 and 2019, respectively. 
For the remaining 18 scam addresses, 17 of them were COVID-19 donations scams. However, we did not identify any obvious scam activities of them before this pandemic. Thus, it is quite possible that they were ordinary addresses before, which were then used in scams during COVID-19.
Most of the scam addresses (86.8\%) are active after 2020.
\textit{It is obvious that, there is a sharp increase in March and April 2020, which is inline with the time of the global outbreak of COVID-19}.
Among these 131 addresses that are new emerging in 2020, the first address that succeeded in receiving crypocurrencies is 0x4e4f4153C6DA6c6df6ecA1f3BF367E5461Ad8F88, which was a giveaway scam that extracted from a Telegram account ``Help\_covid19funds''. 
It received $0.1$ ETH on 7th February and has received $6.5$ ETH (about \$1518.7) in total.

\begin{figure}[t]
\centering
\includegraphics[width=0.45\textwidth]{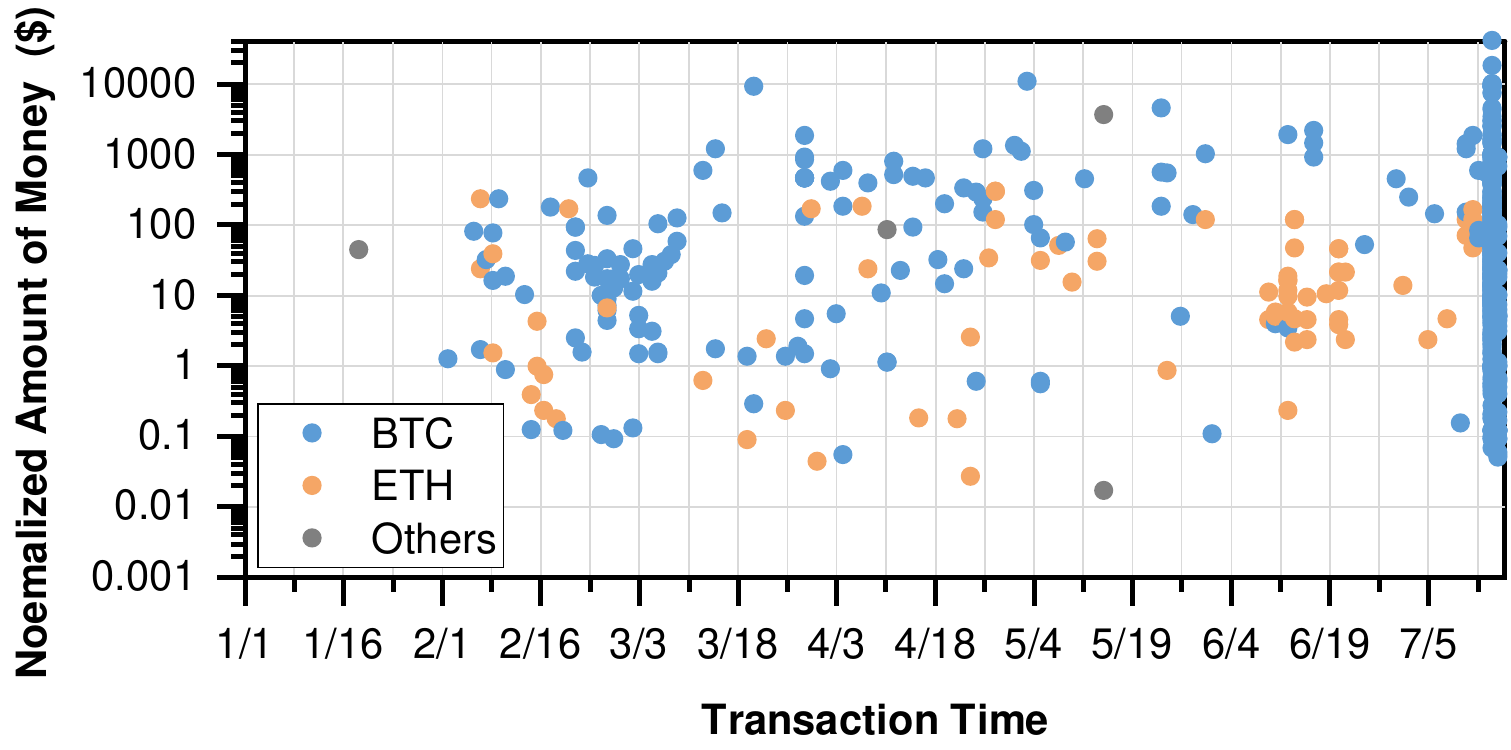}
\vspace{-0.1in}
\caption{The distribution of the time when these scam addresses first received cryptocurrencies. Some scam addresses were active before this pandemic.}
\vspace{-0.1in}
\label{fig:addrscatter}
\end{figure}

\textbf{The Distribution of Transactions.} 
The distribution of the scam addresses' incoming transactions is shown in Figure~\ref{fig:addrscatter}. For the amount of financial losses, most of them are small transactions, i.e.,
\textit{over 90\% of the transactions are below $1,000$ US dollars and 67.6\% of the transactions are below 100 US dollars}.
The largest transaction happened on July 15th, when the address\footnote{bc1qxy2kgdygjrsqtzq2n0yrf2493p83kkfjhx0wlh} associated with the Twitter hack event received 4.56 BTC (roughly $41,642$ US dollars). 
As to the transaction time, first transaction to the scam address happened on January 19th, 2020, and the donation scam address 0x376624e29f8c52b0181bdd794c76fd1058963334 received $44.5$ USDT. 
%Since these address is created before 2020 and has many addresses before, it is highly likely another themed scam carried by the owner of this address. 
The Twitter hack campaign on July 15th accounts for over half of incoming transactions in our dataset, suggesting its great impact on the blockchain community.

% \begin{figure}[t]
% \centering
% \includegraphics[width=0.45\textwidth]{images/4_0_4-moneycdf.pdf}
% \caption{The normalized amount of money distribution of BTC and ETH addresses' transactions.}
% \label{fig:moneycdf}
% \end{figure}

\begin{figure}[t]
\centering
\includegraphics[width=0.45\textwidth]{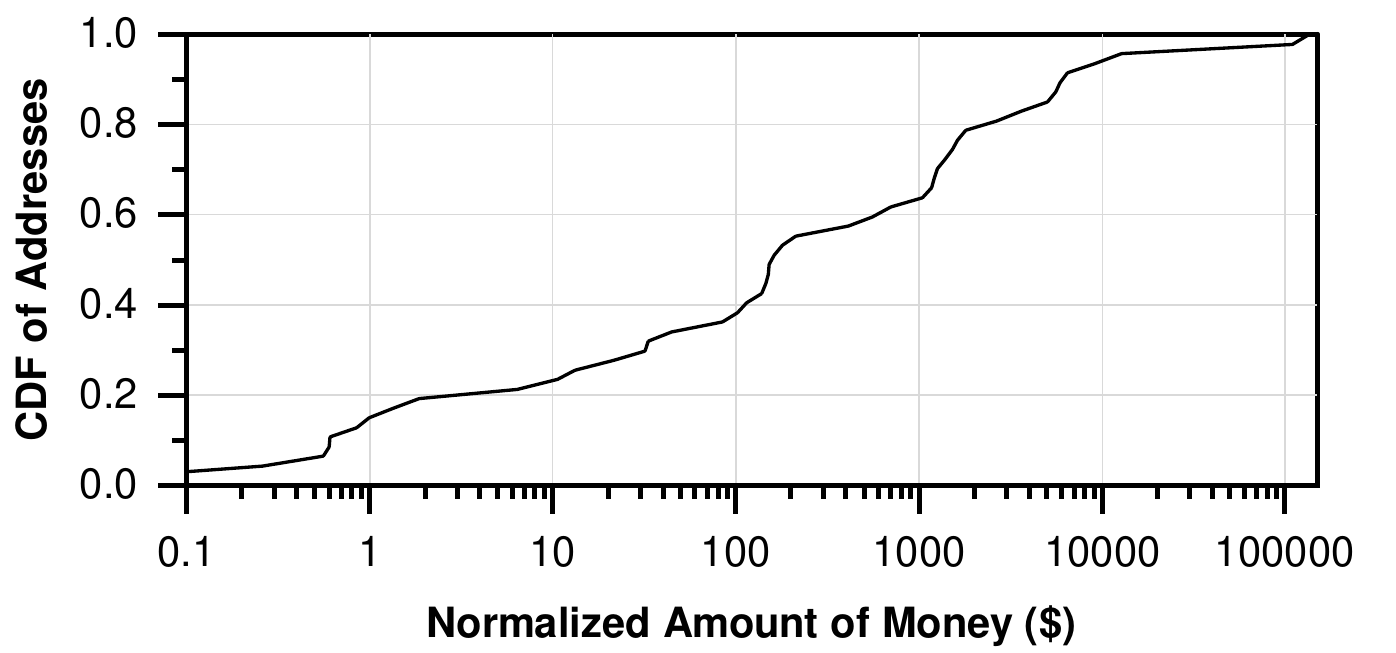}
\caption{The normalized amount of money distribution of BTC and ETH addresses' transactions.}
\label{fig:moneycdf}
\end{figure}

\textbf{The Most Profitable Addresses.} 
On average, each active scam address has received $13.3$ transactions. 
The most active address bc1qxy2kgdygjrsqtzq2n0yrf2493p83kkfjhx0wlh has received 321 transactions. We further analyze the distribution of money each scam address received, as shown in Figure~\ref{fig:moneycdf}. It can be observed that over 61.7\% of the active addresses have received less than \$$1,000$ equivalent cryptocurrencies. 
Table~\ref{tab:topaddr} shows the top-5 most profitable addresses in our dataset. Notably, three of them belong to the Twitter hack related addresses. The largest one has received 14.75 Bitcoins, which is equivalent to roughly \$$134,781.2$.

% \usepackage{booktabs}
\begin{table*}[t]
\centering
\caption{The top 5 profitable addresses in the dataset.}
\resizebox{\linewidth}{!}{
\begin{tabular}{@{}ccccrr@{}}
\toprule
Address                                    & Description                    & Scam Category  & \# of Incoming Transactons & \# of Cryptos Received & Est. Value (\$) \\ \midrule
1Ai52Uw6usjhpcDrwSmkUvjuqLpcznUuyF         & A Twitter hack address         & Giveaway Scams & 39                         & 14.75BTC                 & 134,781.2        \\
bc1qxy2kgdygjrsqtzq2n0yrf2493p83kkfjhx0wlh & A Twitter hack address         & Giveaway Scams & 321                        & 12.02BTC                  & 109,799.9        \\
182P8T7MM9assMo7V9YpqZ925yJX8HZurL & A campaign posing as   Bill Gates' fundatoin & Giveaway Scams & 23 & 1.4BTC & 12,838.2 \\
142pEjfSBh8cnvE7BMJXMjoSBedJWvRKAG         & A campaign reported by Spam404 & Donation Scams & 1                          & 1.0BTC                      & 9,132.2          \\
1cv4tEUqUY7PciJaJWzFCxRuK75veTQNS          & A Twitter hack account         & Giveaway Scams & 6                          & 0.7BTC                   & 6,440.2          \\ \bottomrule
\end{tabular}
}
\vspace{-0.1in}
\label{tab:topaddr}
\end{table*}

\begin{figure}[h]
\centering
\includegraphics[width=0.45\textwidth]{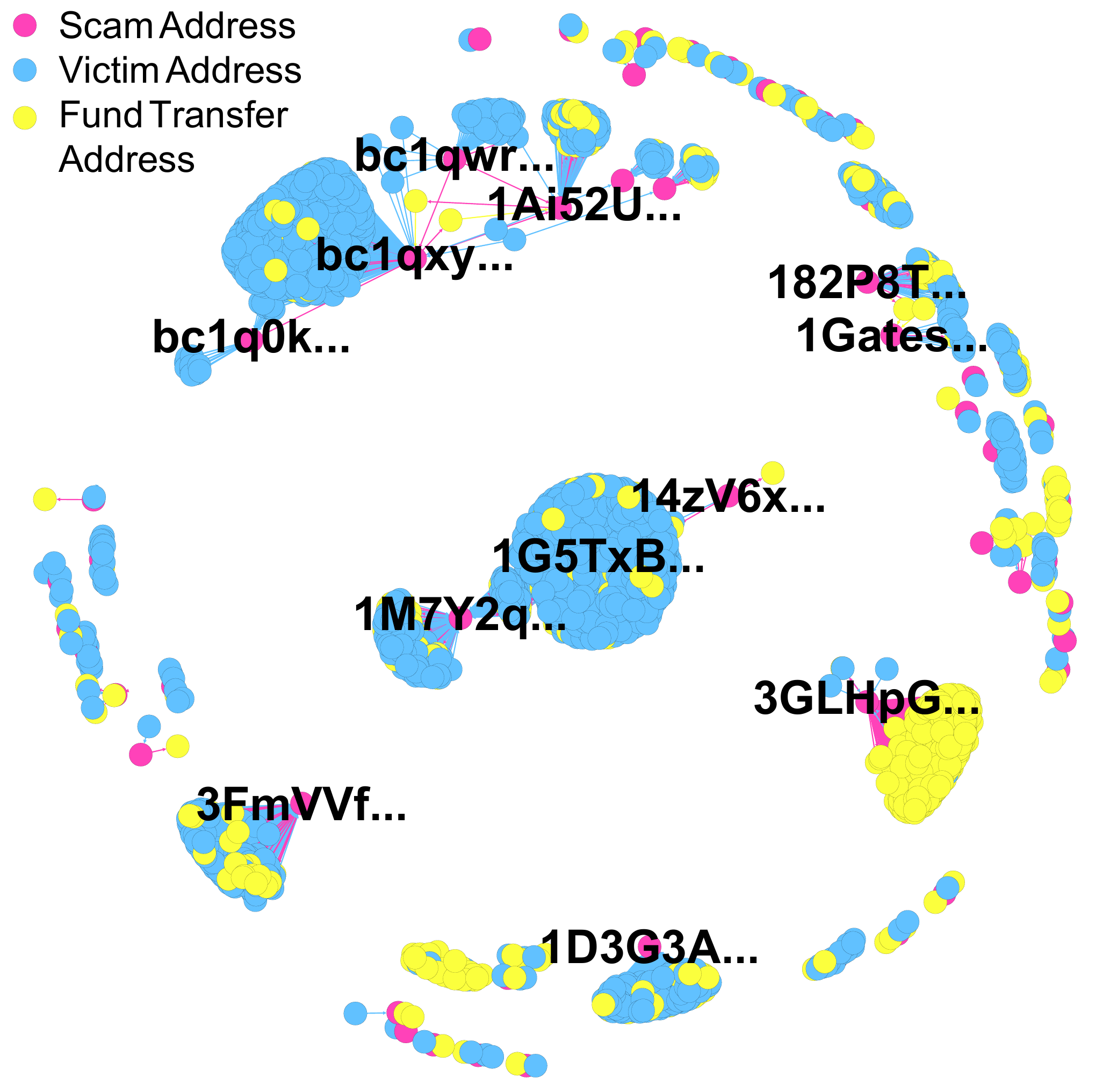}
\caption{The relationship between scam addresses.}
\label{fig:moneyflow}
\end{figure}

\textbf{The Relationship among Scam Addresses.} 
Furthermore, we attempt to investigate relations among scam addresses, i.e., \textit{whether they are controlled by the same malicious campaigns}. For addresses that have ever received transactions on Bitcoin and Ethereum, we analyze their relationship based on the \textit{money flow}, i.e., transactions from one address to another. Figure~\ref{fig:moneyflow} shows the relationship between scam addresses. 
There are 3 types of addresses in the graph: 
1) the labelled \textit{scam addresses} (we exclude silent scam address that have no transaction records);
2) the \textit{victim addresses}, which have ever transferred money to scam addresses but did not receive money from them;
and 3) the \textit{fund transfer addresses}, which are served as the \textit{money laundering channels}, i.e.,
receive money from scam addresses and help the attackers transfer the money they have scammed. 
%%Note that we consider BTC's change address as fund transfer addresses.

There are $56$ scam addresses, $1,741$ victim addresses and $600$ fund transfer addresses shown on Figure~\ref{fig:moneyflow}. 
On average, each BTC scam address is connected to $37.1$ victim addresses and each ETH scam address is connected to $6.5$ victim addresses.
\textit{Interestingly, some scam addresses are clustered into the same group.}
Here, we perform the connected component analysis. As long as there are paths between two scam addresses, we will cluster them together. Note that, we have excluded the impacts introduced by exchange addresses, as different scam addresses can exploit the same exchange address for money laundering. 
As a result, 9 scam addresses are clustered into 3 clusters.
%%\haoyu{can we say how many groups we identified? correspond to how many scam addresses?} \pengcheng{3 groups (2 are reported groups) with 9 addresses, and there are another 2 groups that are connected by exchange addresses.}
For example, the address cluster centred on address bc1qwr30ddc04zqp878c0evdrqfx564mmf0dy2w39l is the Twitter hack address group. They have connected to 533 victim addresses and 38 transfer addresses. Most of the money they received was finally transferred to 1Ai52Uw6usjhpcDrwSmkUvjuqLpcznUuyF, another scam address in the group, and was then transferred to multiple addresses for money laundering. 
We also find a new scam campaign based the relationship of the addresses. They are 182P8T7MM9assMo7V9YpqZ925yJX8HZurL and 1Gatesk17u25gLEk4JNYMDTg8WkCmLpn47, which are connected by 2 fund transfer addresses. It is interesting to see that they both carried out giveaway scams and used similar domain names (\texttt{gatesbtc.live} and \texttt{gatesmicrosoft.tech}) that impersonated as Bill Gates' foundation.

\subsection{COVID-19 Token Scam}
\subsubsection{The Evolution of Tokens} 
\begin{figure}[t]
\centering
\includegraphics[width=0.45\textwidth]{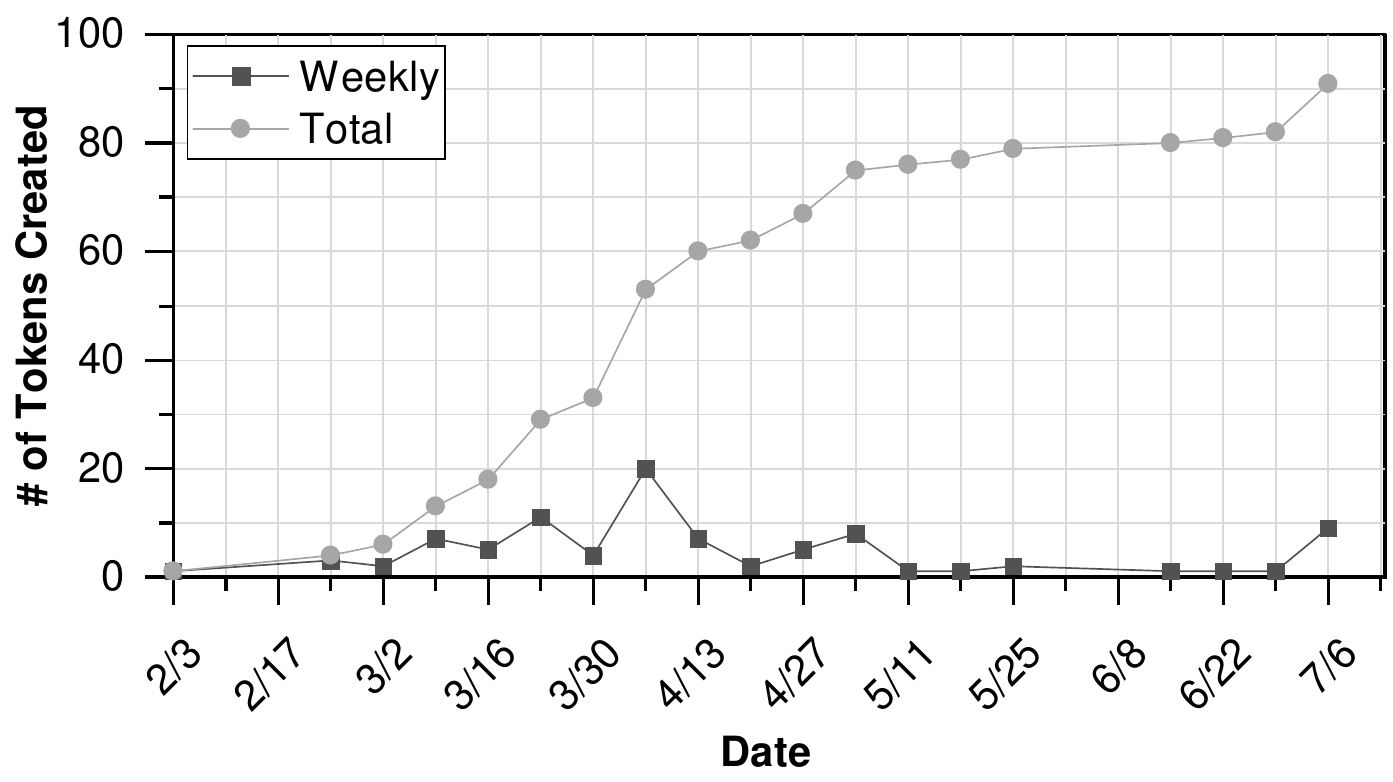}
\vspace{-0.1in}
\caption{Creation time of COVID-19 themed scam tokens.}
\vspace{-0.1in}
\label{fig:tokenonline}
\end{figure}

There are 91 COVID-19 scam tokens in our dataset.
The distribution of their creation time is shown in Figure~\ref{fig:tokenonline}. 
After 2 test tokens created%\footnote{0x0b2aec1a17385727752749d5b47754b433e8c639 on Jan 31st and 0x5567Bb035038cD291d0C88350cF8E3CF23455700 on Feb 3rd, they have no holders and little transfers.}
, the first CoronaCoin\footnote{Address:0x10Ef64cb79Fd4d75d4Aa7e8502d95C42124e434b} was officially released on 4th February 2020, and a large number of tokens are created after that. Notably, when the coronavirus raised global concern in March, there are 31 (34.1\%) newly emerging tokens created.

\subsubsection{Current Status of Tokens} 
\begin{figure}[t]
\centering
\includegraphics[width=0.48\textwidth]{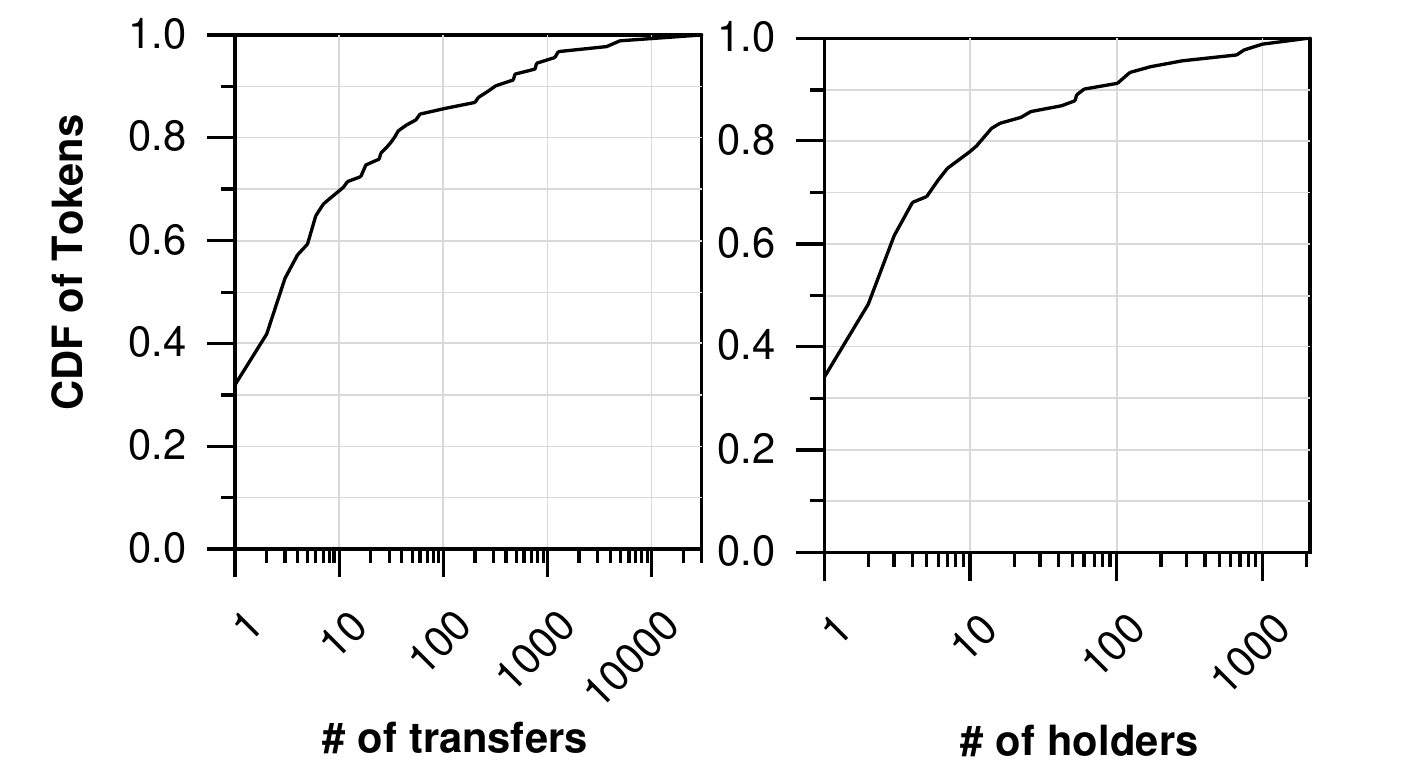}
\vspace{-0.1in}
\caption{The number of holders and transfers of scam tokens.}
\vspace{-0.1in}
\label{fig:tokentransferholder}
\end{figure}

The number of token's holders and transfers can reflect the activity of a token to a certain extent. Figure~\ref{fig:tokentransferholder} shows the current\footnote{We crawled the data until block $10,473,613$, i.e. UTC 2020-07-16 11:59:59 PM.} status of these tokens. 84.62\% of the tokens have less than 60 transfers and 86.81\% of the tokens have less than 50 holders, indicating that most scam tokens do not infect too many victims. Among them, 5 tokens have more than 200 holders (see Table~\ref{tab:toptoken}). Note that, the first 3 tokens have the same name `CoronaCoin', we add the suffix based on their online time to distinguish them.

% \usepackage{booktabs}
\begin{table}[t]
\caption{Top 5 tokens with the most number of holders.}
\resizebox{\linewidth}{!}{
\begin{tabular}{@{}lllll@{}}
\toprule
Token Name &
  Token Address &
  \begin{tabular}[c]{@{}l@{}}\# of \\ Holders\end{tabular} &
  \begin{tabular}[c]{@{}l@{}}\# of \\ Transfers\end{tabular} &
  \begin{tabular}[c]{@{}l@{}}Online \\ Date\end{tabular} \\ \midrule
CoronaCoin \#1    & 0x10Ef64cb79Fd4d75d4Aa7e8502d95C42124e434b & 2016 & 29896 & Feb 4th  \\
CoronaCoin \#3    & 0x0c2c5E2b677dEa43025B5DA5061fEcE445f0295B & 998 & 3666  & Apr 3rd  \\
CoronaCoin \#2    & 0xb80112E516DAbcaC6Ab4665f1BD650996403156C & 740  & 4906  & Mar 30th \\
COVID19           & 0x26ADdc98fF2321A91a776E52be171a963720A42C & 659  & 1270  & Apr 2nd  \\
Corona Coin & 0x170467C28C4BF99f2D3840E730498F730a526Da2 & 281  & 320   & Apr 5th  \\ \bottomrule
\end{tabular}
}
\label{tab:toptoken}
\end{table}

%Top 5 tokens that have most holders is shown on Table~\ref{}. xx\% of the tokens have less than xx holders while xxx tokens have more then 500 holders. The most popular token till now is the CoronaCoin we mentioned in the former section,which has xxx holders. The interesting truth is that CoronaCoin #2 is the fork of CoronaCoin #1 after the lead developer comprised the donation wallet and CoronanToken is newly created after the 3rd exit scams of the CoronaCoin developer team. And all this 3 tokens still have new transactions till the time we research\footnote{}.

\subsubsection{Price and Volume} 
\begin{figure}[t]
\centering
\includegraphics[width=0.45\textwidth]{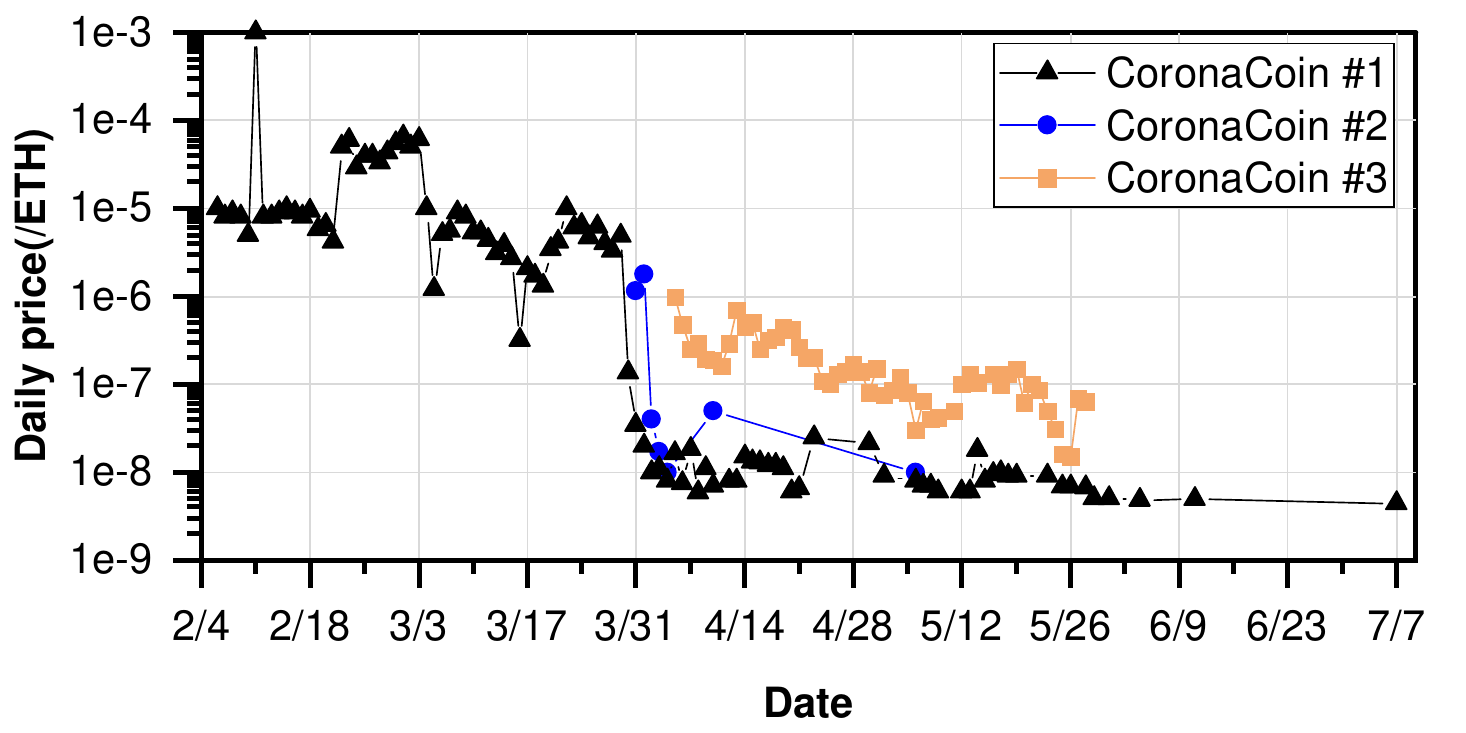}
\vspace{-0.1in}
\caption{Daily closing price of the top 3 tokens ranked by average volume.}
\vspace{-0.1in}
\label{fig:tokenprice}
\end{figure}

\begin{figure}[t]
\centering
\includegraphics[width=0.45\textwidth]{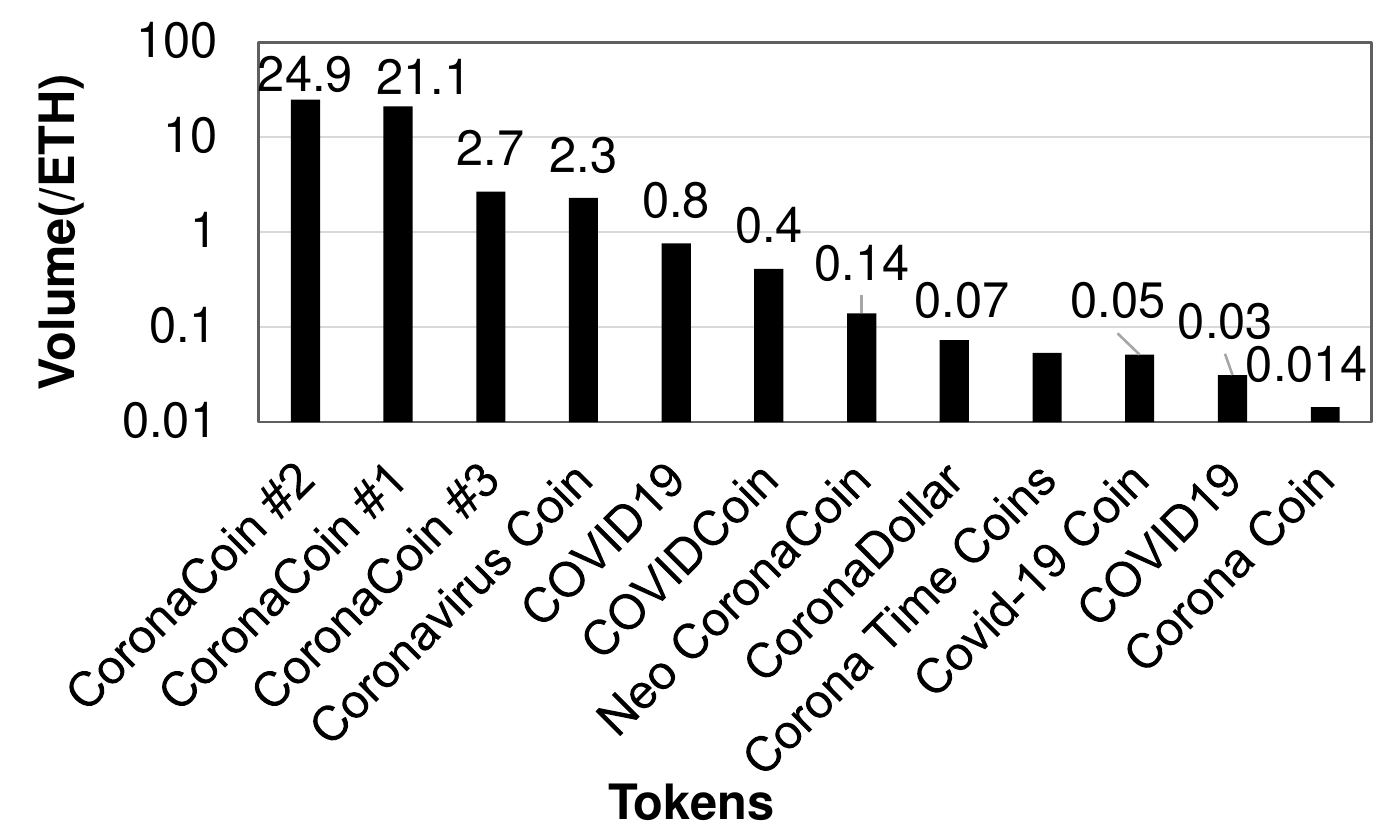}
\vspace{-0.1in}
\caption{Daily average volume of tokens on Saturn Network.}
\vspace{-0.1in}
\label{fig:tokenavgvolume}
\end{figure}

Among these tokens, several tokens are listed on some decentralised cryptocurrency exchanges (DEX), so that we can fetch the daily price and volume of them. 
We choose the price and volume from Saturn Network Exchange\footnote{https://www.saturn.network/} because most available tokens (16) are listed there. Their daily average volume is shown on Figure~\ref{fig:tokenavgvolume}\footnote{Note that due to space limitation, we only list the top-12 tokens based on their volumes.}. Figure~\ref{fig:tokenprice} shows the daily closing price of top 3 tokens that have the most average volume. The highest price $0.001$ ETH of CoronaCoin \#1 occurred on Feb 11th, but considering its normal volume that day, we believe it is just a single case that the transaction is likely to be initiated by the founders of CoronaCoin \#1. Their sole purpose is to increase the price of this token. After that, the price of CoronaCoin \#1 rises due to increasing attention. On March 4th, a developer of the founding team made an exit scam by dumping all the tokens he had, which makes the price decreased sharply. The CoronaCoin \#2 and \#3 both created as forks of their former version because of exit scams, and that is why the price curve of CoronaCoin \#1 and \#2 dropped rapidly in late March and early April.

\iffalse
\begin{table*}[t]
\centering
\caption{The top 5 valuable tokens. }
\resizebox{\linewidth}{!}{
\begin{tabular}{@{}ccccccc@{}}
\toprule
Token name & Address & \# of Transfers & \# of Holders & Current token value (\$) & Scammed Amount (\$) & Est. Total Value \\ \midrule
CoronaCoin \#1 & 0x10Ef64cb79Fd4d75d4Aa7e8502d95C42124e434b & 29896 & 2,016 & 33.5   & 18,310.4 & 18,343.8 \\
CoronaCoin \#2 & 0xb80112E516DAbcaC6Ab4665f1BD650996403156C & 4906  & 740  & 76.0   & 2336.4  & 2,412.4  \\
COVID19        & 0x6b466b0232640382950c45440ea5b630744eca99 & 756   & 53   & 1,178.0 & 0.0     & 1,178.0  \\
CoronaCoin \#3 & 0x0c2c5E2b677dEa43025B5DA5061fEcE445f0295B & 3666  & 998  & 491.4  & 0.0     & 491.4   \\
COVIDCoin      & 0x9947A675Cb4D4A19e020E1DD035955c0150b1e5e & 465   & 52   & 203.2  & 0.0     & 203.2   \\ \bottomrule
\end{tabular}
}
\label{tab:tokenvalue}
\end{table*}
\fi

\begin{table}[t]
\centering
\caption{Top-5 addresses involved in giveaway scams.}
\resizebox{\linewidth}{!}{

\begin{tabular}{@{}cccc@{}}
\toprule
  Address &
  \begin{tabular}[c]{@{}c@{}}\# of \\ Incoming \\ Transactions\end{tabular} &
  \begin{tabular}[c]{@{}c@{}}\# of \\ Cryptos \\ Received\end{tabular} &
  \begin{tabular}[c]{@{}c@{}}Est. \\ Value (\$)\end{tabular} \\ \midrule
 bc1qxy2kgdygjrsqtzq2n0yrf2493p83kkfjhx0wlh & 321 & 12.02 BTC & 109,799.9 \\
 0xd03dc334fb65cea1b83e654b26515e72694a713f & 41  & 5.92 ETH  & 1,383.2   \\
 1Ai52Uw6usjhpcDrwSmkUvjuqLpcznUuyF         & 39  & 14.75 BTC & 134,781.2 \\
 bc1qwr30ddc04zqp878c0evdrqfx564mmf0dy2w39l & 36  & 0.55 BTC  & 5,050.4   \\
 182P8T7MM9assMo7V9YpqZ925yJX8HZurL         & 23  & 1.41 BTC  & 12,838.2  \\ \bottomrule
\end{tabular}
}
\label{tab:addrgiveaway}
\end{table}

\subsubsection{ICO Scam and Exit Scam} 
%%According to the data we got through our detection method, including the 3 exit scams we mentioned in the former section, we find 7 scams about ICO scams and exit scams. 
By manually investigation, we found 7 out of the 91 scam tokens have behaviors like \textit{ICO scam} and \textit{exit scam}. For the CoronaCoin \#1, it has two evident exit scams. 
One is on March 4th, when a developer dumped 5 million tokens from the developers' wallet, which profited him $14.94$ ETH ( \$$3,490.6$)\footnote{The actor's address:0x1865E7b66c3996Bd55e7dD88a16B897f4123D2A7}. On March 30th, the lead developer compromised the wallet\footnote{Address:0xf19afb42e574831776e7af0898f05c1b306bf3b7} that originally intended to donate to the Red Cross, and he got $63.43$ ETH (about \$$14,819.8$). The rest of the team migrated the project to the CoronaCoin \#2, and shortly after that, the third dump happened where the malicious actors\footnote{Address:0x4121CB0d7AA53AAbc7596c9cAfaa02B1863a2aC7} got 10 ETH (about \$$2,336.4$).
Similar Pump-and-dump scams are found in VaccinaCoin\footnote{Address:0x567d297d0cbb66195b268162a4547f220ef49c51, which is found on BitcoinTalk and claims to help COVID-19 vaccine production.} and COVID19 token\footnote{Address:0x6b466b0232640382950c45440ea5b630744eca99}.
%were identified by people due to their untenable description in advance, but they are stll listed on exchanges and active for some days. 
There are two scam tokens involved in the ICO scams, i.e., they either provide fake information or entice users to buy the token using giveaway-like tricks. Luckily, these ICO scams have not scammed many users yet. For the Corona Virus Coin\footnote{Address:0x49017D1cE3359a3b81AE8417731298126Ff751F1}, it only tricked 1 victim successfully with merely $2.6$ dollar profit. Another one is CoronaAid\footnote{Address:0xd3CD8Ce0c357CAdC16F812884c2dDb89F8A22103}, which has not received money yet by the time of this study.

\begin{figure}[t]
\centering
\includegraphics[width=0.45\textwidth]{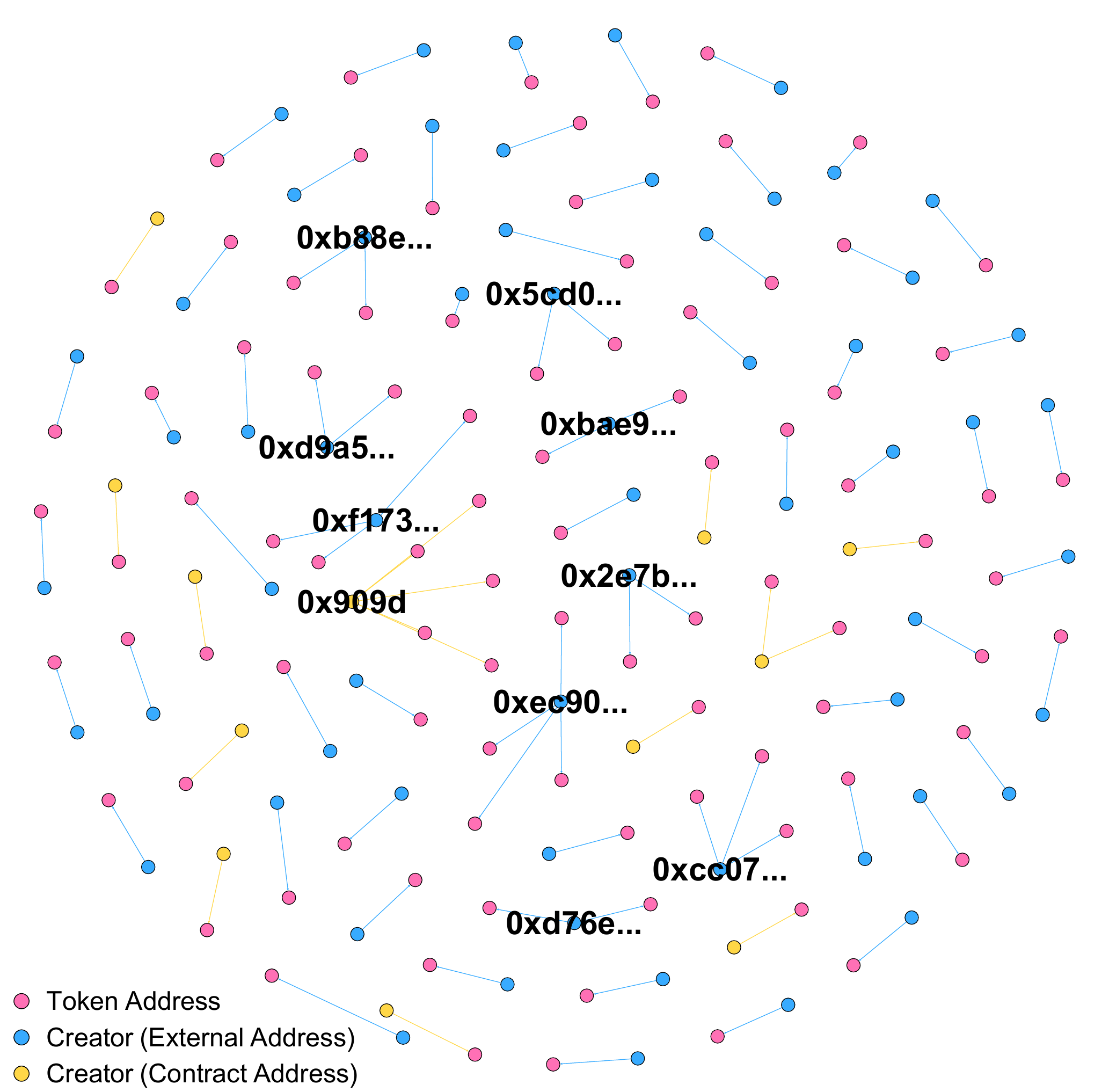}
\caption{The creator graph of scam tokens.}
\vspace{-0.1in}
\label{fig:tokencreator}
\end{figure}

\subsubsection{The Creators of Scam Tokens}
We further analyze the creators of scam tokens, as shown in Figure~\ref{fig:tokencreator}. 
Overall, 73 creators released 91 tokens.
Over 81\% (74) of the tokens are created by external addresses (i.e., by humans), while the remaining tokens (17) are created automatically (i.e., by smart contracts).
We observe that, some creators (9 external addresses and 2 contract addresses) have released more than one COVID-19 scam token.
For example, the address 0xec9005224daa378598aa0ea8f4c656d5a7c6de76 created 4 tokens. After creating 3 tokens that are less popular, this address created the 4th token named ``Corona Coin''\footnote{Address:0x170467C28C4BF99f2D3840E730498F730a526Da2}, which has 281 holders by the time of this study. 
%%Most of the addresses that created multiple tokens are acting like this way. 
Interestingly, 5 scam tokens are created by token template contract MiniMe\footnote{https://github.com/Giveth/minime, address: 0x909d05F384D0663eD4BE59863815aB43b4f347Ec}, which has reduced the efforts of creating scam tokens.

%%Considering that so many scams happened in the field of coronavirus-themed tokens, it is much possible that other valuable tokens are dumped by bad actors. Thus, we estimated these tokens' value by multiplying their close price in the last price record on Saturn Exchange and the sum of tokens that their holders hold\footnote{Note that, we exclude tokens whose holders are under 10, because this kind of tokens are more likely to have unconventional transactions and their price may be amazingly high when calculated using this method. }. The Table~\ref{tab:tokenvalue} below is a table summarizing the top 5 valuable tokens. In total, these tokens' value is $23177.99$ US Dollars.

% \usepackage{booktabs}
% \usepackage{booktabs}
\begin{table}[h]
\centering
\caption{Crypto blackmails related to COVID-19.}
\resizebox{\linewidth}{!}{

\begin{tabular}{@{}ccccc@{}}
\toprule
Address &
  Trick &
  \begin{tabular}[c]{@{}c@{}}\# of \\ Incoming \\ Transactions\end{tabular} &
  \begin{tabular}[c]{@{}c@{}}\# of \\ Received \\ Cryptos\end{tabular} &
  \begin{tabular}[c]{@{}c@{}}Est. \\ Value (\$)\end{tabular} \\ \midrule
1PQMFqfnEtddpPwHyB3E3jWvakDZRmZtCg         & Selling products    & 2 & 0.023 BTC   & 213.0 \\
1NM8LLAGcMHjPZJ5ysUWvNrVjwFtQnYjYe         & Selling products    & 1 & 0.017 BTC   & 151.5 \\
1JoWQUtGcP2yYdQwDcjsUxgrZjntrBqLLQ         & Selling products    & 2 & 0.016 BTC   & 149.9 \\
18P3S6DuNUpW2WLozsrrW6rRd6xh24Rc7N         & Virus spread threat & 1 & 594 Satoshi & 0.1   \\
bc1q9l5tvr322w36ky9ex8kfhzjhffahmqmxg8s3we & Virus spread threat & 0 & 0           & 0     \\
bc1q9tpt4dffddj4dgtt8c8meeghvcqveml8j4n9wz & Virus spread threat & 0 & 0           & 0     \\
bc1qxaujxahx5wvfe4dy4eq5q939yw7ffmj8tctwpj & Virus spread threat & 0 & 0           & 0     \\
bc1q2868603740f4y3yq6x7ksmqq2gyynyeegpvtc4 & Virus spread threat & 0 & 0           & 0     \\
3EWMDjjY3zimYkbNb9od2uxVeuVREcDpGw         & Selling cure        & 0 & 0           & 0     \\ \bottomrule
\end{tabular}
}
\label{tab:blackmail}
\end{table}

% \usepackage{booktabs}
% \usepackage{multirow}
\begin{table*}[h]
\centering
\caption{Crypto malware with blockchain addresses.}
%%\vspace{-0.1in}
\resizebox{\linewidth}{!}{
\begin{tabular}{@{}ccccc@{}}
\toprule
App Name & Source  & MD5                              & Address                            & Total Received(\$) \\ \midrule
\multirow{4}{*}{Coronavirus Tracker} &
  \multirow{4}{*}{coronavirusapp.site} &
  \multirow{4}{*}{\begin{tabular}[c]{@{}c@{}}d1d417235616e4a05096319bb4875f5,   \\ 1602c0258f39b2b032edd7d6160befe7, \\ 339de9104962d6c8cfd6d8b4a68bcb4d,   \\ 69a6b43b5f63030938c578eec05993eb\end{tabular}} &
  \multirow{4}{*}{18SykfkAPEhoxtBVGgvSLHvC6Lz8bxm3rU} &
  \multirow{4}{*}{\begin{tabular}[c]{@{}c@{}}98.4\end{tabular}} \\
         &         &                                  &                                    &                    \\
         &         &                                  &                                    &                    \\
         &         &                                  &                                    &                    \\\hline
Covid19  & Koodous & 362dac3f2838d2bf60c5c54cc6d34c80 & 3HEtt3VHoAj18rDbJoYv6mbojBK4DY9zyu & 0                  \\\hline
WSHSetup.exe &
  www.wisecleaner.best &
  ec517204fbcf7a980d137b116afa946d &
  \begin{tabular}[c]{@{}c@{}}bc1qkk6nwhsxvtp2akunhkke3tjcy2wv2zkk00xa3j,\\ bc1q8r42fm7kwg68dts3w70qah79n5emt5m76rus5u\end{tabular} &
  0 \\ \bottomrule
\end{tabular}
}
\vspace{-0.1in}
\label{tab:ransomware}
\end{table*}

\subsection{COVID-19 Giveaway Scam}
In total, we find 19 giveaway scams, to which 5 domains and 10 social accounts are related. Among them, 14 of 21 extracted addresses have succeeded in receiving cryptocurrencies in 2020. Table~\ref{tab:addrgiveaway} shows top-5 active addresses. The most popular address is \textit{bc1qxy2kgdygjrsqtzq2n0yrf2493p83kkfjhx0wlh}, which received 321 transactions. This address is a Twitter hack address, which is not used until July 15th. Besides the Twitter hack addresses, the most active address in Ethereum is 0xd03dc334fb65cea1b83e654b26515e72694a713f, whose attackers pretend to be \textit{Vitalik Buterin}, a co-founder of Ethereum. The attacker gets \$$1383.2$ equivalent profit. It can be observed that COVID-19 themed giveaway scams impersonating famous persons or organizations are more effective.

\subsection{COVID-19 Crypto Blackmail}
\label{subsec:covid19-balckmail}
We have collected 9 addresses related to blackmails, all of which are Bitcoin addresses. They are totally reported 23 times on BitcoinAbuse by users.
The profit they made is shown in Table~\ref{tab:blackmail}. To our surprise, attackers do not make too much profit through these tricks. The most profitable address has received 0.023 BTC, which claims to sell products during the pandemic. 6 of the scam addresses do not even receive any money. 
It might because people's awareness of such scams has increased, while scaring people with virus by email is difficult to cheat people.
%%\haoyu{can also summarize the tricks they used in the Email. may use a table here.}

\begin{table}[h]
\centering
\caption{Top 5 most profitable donation scam addresses.}
\vspace{-0.1in}
\resizebox{\linewidth}{!}{

\begin{tabular}{@{}cccc@{}}
\toprule
  Address &
  \begin{tabular}[c]{@{}c@{}}\# of \\ Incoming \\ Transactions\end{tabular} &
  \begin{tabular}[c]{@{}c@{}}\# of \\ Cryptos \\ Received\end{tabular} &
  \begin{tabular}[c]{@{}c@{}}Est. \\ Value (\$)\end{tabular} \\ \midrule
 142pEjfSBh8cnvE7BMJXMjoSBedJWvRKAG & 1  & 1 BTC   & 9,132.2    \\
 TVBkqA4BuAoy8p3tNXMvHeC45BuC7PHJw3 & 2  & 211,254.33 TRX & 3,621.7 \\
 3GLHpGWbBi1cAWUbn4mSx5RJygDmA6d89o & 3  & 0.29 BTC   & 2,661.3 \\
 1G5TxBiZ8JDYNgUnjsY8KoAQxq2Sj6dKby & 45 & 0.19 BTC   & 1,805.2 \\
 15f7eGQt2CLJJB86jt5whrZQ5RRERCT574 & 5  & 0.14 BTC   & 1,257.4 \\ \bottomrule
\end{tabular}
}
\label{tab:donationaddr}
\end{table}

\subsection{COVID-19 Crypto Malware}
As to COVID-19 themed crypto malware, we have identified 5 Android ransomware and 9 Windows malware binaries (4 labeled as Ransomware and 5 labeled as Coinminer). 
By manually investigation, we found 4 addresses in them (1 address shared by 4 Android apps and 2 addresses embedded in 1 windows ransomware). 
Their profits are shown in Table~\ref{tab:ransomware}.
The only address which made a profit is related to the domain \texttt{http://coronavirusapp.site} and its distributed apps are disguised as coronavirus tracker apps.

\subsection{COVID-19 Ponzi Scheme}
\begin{figure}[t]
\centering
\includegraphics[width=0.45\textwidth]{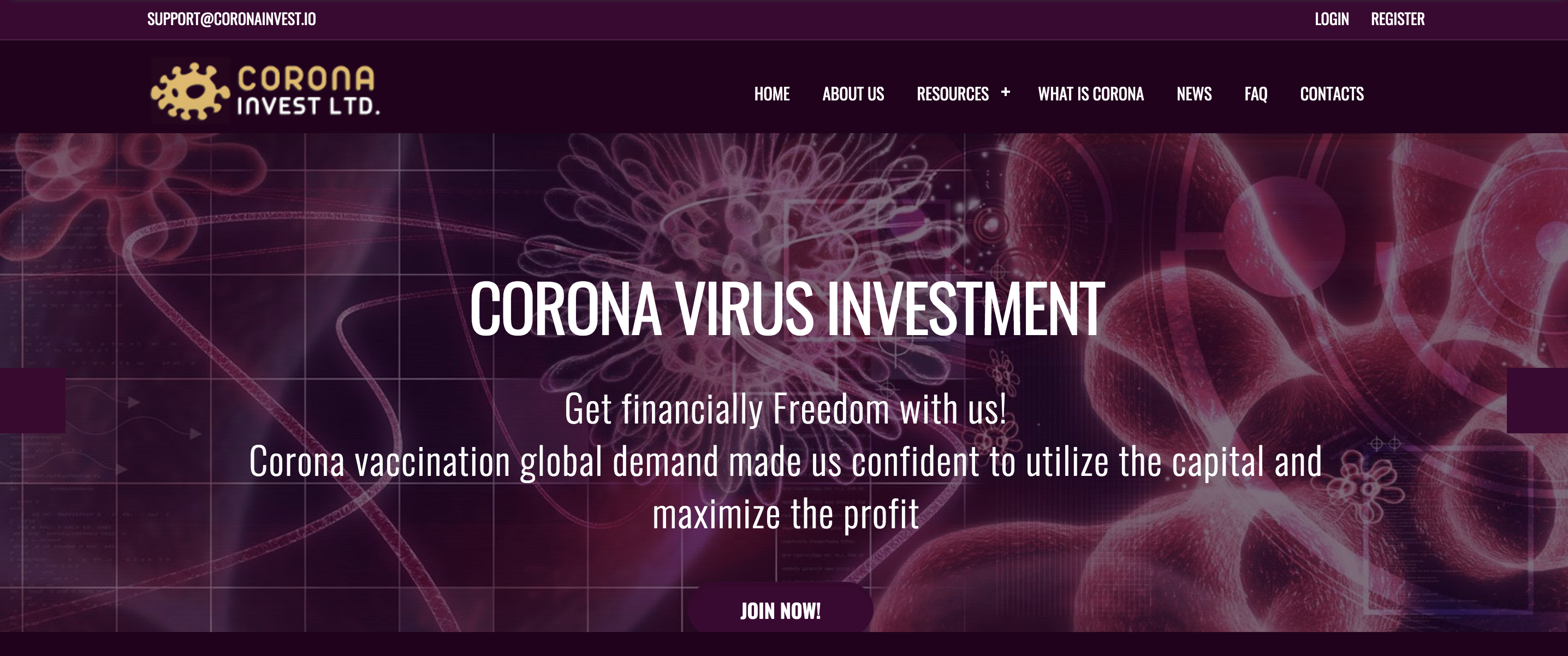}
\caption{An example of a scam description on https://coronainvest.io/.}
\label{fig:ponzi2}
\end{figure}

We find 9 websites that are suspicious to carry out Ponzi schemes. For example, they claimed that they will invest in vaccine of COVID-19 and make the investors get financial freedom, as shown in Figure~\ref{fig:ponzi2}. Table~\ref{tab:ponziweb} lists the 9 websites. We can see that, most of them contain the keyword ``invest'' in their domain names, and the descriptions shown on their websites indicate they are highly suspicious to be scams.
However, by the time of our study, none of them can function well or they are just unreachable during our manually analysis. 
Thus, we cannot get their scam blockchain addresses, so that we cannot estimate their impacts and the number of victims.

\begin{table}[h]
\centering
\caption{A list of COVID-19 themed Ponzi scheme domains.}
\vspace{-0.1in}
\begin{tabular}{@{}|c|c|@{}}
\toprule
\multicolumn{2}{|c|}{Ponzi Scheme Websites}                        \\ \midrule
coronainvest.io       & covidinvestmenthelp.com                    \\
coronabit.cc          & covid19invest.com                          \\
coronafeverinvest.com & covid-19investmentchallenge.azonixtech.com \\
covid-invest.tk       & covid-invest.laber.ru                      \\
coronacoin.xyz        &                                            \\ \bottomrule
\end{tabular}
\vspace{-0.1in}
\label{tab:ponziweb}
\end{table}

%%\haoyu{can you show the case of Telegram account here?}
%%\haoyu{can you list all the websites or account in a table?}

\subsection{COVID-19 Fake Crypto Donation}
We have identified 76 blockchain donation scam addresses, which are related to 23 domains and 30 social accounts. 
Table~\ref{tab:donationaddr} shows the top 5 most profitable addresses. 
The most profitable address 142pEjfSBh8cnvE7BMJXMjoSBedJWvRKAG received 1 BTC on March 22nd.
%%, whose scam campaign was reported by Spam404\footnote{https://otx.alienvault.com/pulse/5e7f99a0808c6d74674980ca}. 
In total, 28 of 76 addresses have made a profit of $21788.9$ US dollars, which accounts for the third largest scam category by profit in our dataset. Because of donation's close connection with this pandemic, such a serious condition deserves our vigilance.

%% file: Discussion.tex
\section{Discussion}
\label{sec:discussion}

\subsection{Implication}
Our observations are of key importance to the stakeholders in the blockchain ecosystem and the researchers who are interested in COVID-19 themed cybersecurity. On one hand, considering the prevalent COVID-19 themed crypto scams in the wild and their great impacts, it is urgent for the community to detect these scams and eliminate possible losses. 
On the other hand, it suggests that the attackers are taking advantage of this kind of public event to perform cyber-attacks.
Although the six types of scams studied in this paper follow the similar behavior patterns as other non-Coronavirus scams suggested in previous work, they have adopted a number of social engineering techniques to deceive users, and similar techniques can be easily adopted to other social events or domains. Thus, the governance of the cyberspace on social events should be improved. For example,
paying special attentions to the newly released domains/tokens/social network accounts that related to the public events can help us identify these scams timely.
Furthermore, identifying the distribution channels (e.g., Twitter, Telegram, and malicious domains, etc.) of these scams can greatly reduce the impacts.

\subsection{Limitation}
%%To the best of our knowledge, our paper is the first systematic study of COVID-19 themed cryptocurrency scams. 
Our study carries several limitations.
First, the taxonomy of scams might not be complete.
As the taxonomy is summarized based on public known scams reported by users in the wild, it is quite possible that there are new tricks we did not identified. Nevertheless, we believe we have covered most kinds of the COVID-19 themed crypto scams.
Second, as suggested in Section~\ref{subsec:distribution}, a few scam addresses were active before 2020. Although we only count their transactions after the break of COVID-19 (only 7 of them have transactions in 2020), it is hard to differentiate how many of them are actually involved in COVID-19 scams. Nevertheless, we found that most of them have very few transactions before 2020, and there is usually a quiet period till their rejuvenation in COVID-19. As a result, we believe most of the revenue should be credited to COVID-19. 
Furthermore, it is quite possible that one scam address can be used to fulfill more than one type of scam activities, while is hard to us to differentiate.
At last, this work relies on some manually efforts to identify the unrevealed scams, which might not be scalable. 
Although we have tried our best to reduce the efforts, i.e., by using VirusTotal to label the suspicious scams first, and applying heuristics to mark COVID-19 related scams, we admit that some advanced techniques (e.g., machine learning techniques) can be implemented to identify and classify the scams in the future.
However, while the study is by no means comprehensive, we are able to provide a lower bound estimate of the prevalence
and criminal profits associated with these COVID-19 scams.

\section{Conclusion}
\label{sec:conclusion}

%A number of malicious campaigns have started capitalizing the topic of COVID-19. 
In this paper, we take the first step to characterize COVID-19 themed cryptocurrency scams.
We investigated six types of cryptocurrency scams related to COVID-19 by collecting existing scam reports and detecting the unrevealed ones. 
Specifically, we revealed how the scams work, measured their prevalence in the wild, studied their evolution and characterized their impacts.
Besides, we released the labeled scam dataset to the research community to help fight against the COVID-19 attacks in cyberspace.